\documentclass[prb,superscriptaddress,twocolumn,aps,showpacs,amsmath,amssymb,floatfix]{revtex4-2}

\usepackage{color}
\usepackage{soul}
\usepackage{comment}
\usepackage[colorlinks]{hyperref}
\usepackage{graphicx}
\usepackage{braket}
\usepackage{MnSymbol}
\usepackage{amsmath,amssymb,bbold,bm}
\usepackage[normalem]{ulem}

\renewcommand{\v}[1]{\textbf{\textit{#1}}}
\newcommand{\abs}[1]{\left\vert #1 \right\vert}
\newcommand{\eps}{\epsilon}
\newcommand{\dg}{^{\dagger}}
\newcommand{\dn}{^{\vphantom{\dagger}}}

\newcommand{\matc}[2]{\left(\bmx{#1}#2\emx\right)}
\newcommand{\tr}{\text{tr}}
\newcommand{\Tr}{\text{Tr}}
\newcommand{\bmx}{\begin{array}}
\newcommand{\emx}{\end{array}}
\newcommand{\mat}[1]{\left(\bmx{cc}#1\emx\right)}
\newcommand{\be}{\begin{equation}}
\newcommand{\ee}{\end{equation}}
\newcommand{\bea}{\begin{eqnarray}}
\newcommand{\eea}{\end{eqnarray}}

\newcommand{\sgn}[1]{\text{sgn}({#1})}

\def\XXint#1#2#3{{\setbox0=\hbox{$#1{#2#3}{\int}$}
     \vcenter{\hbox{$#2#3$}}\kern-.5\wd0}}

\newcommand{\EJK}[1]{{\color{black} #1}}
\newcommand{\yk}[1]{{\color{black} #1}}
\newcommand{\red}[1]{\textcolor{black}{#1}}

\begin{document}
\setstcolor{cyan}
\title{Frustrated Kondo impurity \yk{triangle}: a simple model of deconfinement}

\author{Elio J. K\"onig}
\affiliation{Department of Physics and Astronomy, Center for Materials Theory, Rutgers University, Piscataway, NJ 08854 USA}
\affiliation{Max Planck Institute for Solid State Research, Heisenbergstr. 1, 70569 Stuttgart, Germany.}
\author{Piers Coleman}
\affiliation{Department of Physics and Astronomy, Center for Materials Theory, Rutgers University, Piscataway, NJ 08854 USA}
\affiliation{Department of Physics, Royal Holloway, University of London, Egham, Surrey TW20 0EX, UK}
\author{Yashar Komijani}
\affiliation{Department of Physics and Astronomy, Center for Materials Theory, Rutgers University, Piscataway, NJ 08854 USA}
\affiliation{Department of Physics, University of Cincinnati, Cincinnati, Ohio 45221-0011, USA}
\date{\today}

\begin{abstract}
The concepts of deconfinement and topological order are of great current interest 
for \red{quantum information science and for} our 
understanding
of quantum materials. 
Here, we introduce a 
simple model of three antiferromagnetically coupled Kondo
impurities, a ``Kondo \yk{triangle}'', which can be used to further \yk{extend} the
application of these concepts to electronic systems. 
We show that by tuning the magnetic frustration, the Kondo \yk{triangle} undergoes a
quantum phase transition between two phases of unbroken
symmetry, signaling a phase transition beyond the Landau paradigm. \red{We demonstrate that t}he frustrated ``spin liquid'' phase \red{is described by a three-channel Kondo fix\yk{ed} point and thus displays}
an irrational ground state
degeneracy. \red{Using an Abrikosov pseudofermion representation this quantum state is categorized \yk{by an emergent U(1) gauge field} and its projective symmetry group. The gauge theory is deconfining in the sense that a miniature Wilson loop orders and that topological defects \yk{(instantons in the gauge field)} are expelled. 	This phase persists in presence of moderate Kondo screening until proliferation of topological defects lead to a quantum \red{phase} transition \red{to an unfrustrated Fermi liquid} phase. \red{Based on this evidence, we propose that three-channel Kondo phase displays topological order in a similar sense as gapless spin liquids.}}
\end{abstract}
\date{\today}

\maketitle
%

\section{Introduction}

Recent discoveries in quantum
materials \yk{have urged us to generalize Landau's notion of broken symmetry by
introducing new classes of quantum order without symmetry breaking. The prime example is quantum magnetism in which} strong frustration can give rise to
\red{quantum} spin liquids \red{(QSLs)}\cite{Savary2016,Zhou17} with fractionalized quasi-particles, new patterns of long-range entanglement {and} topological order \cite{Wen17}. Similar physics occurs at continuous phase transitions between ordered phases with different symmetries
which require a fractionalized description (``deconfined criticality'')~
\cite{Senthil04}.

These ideas \yk{are} of particular relevance to 
doped \red{QSLs} in the vicinity of
Mott-delocalization, a topic  
of potential  importance for cuprate~\cite{Anderson1987}, organic
salts ~\cite{KurosakiSaito2005,YamashitaKanoda2008} and
iron-based~\cite{ColemanKoenig2019} high-T$_c$ superconductors.
A closely related topic, is the interaction of electrons and spin
liquids via a Kondo interaction, as in 
geometrically frustrated heavy fermion compounds, 
(e.g.~CePdAl~\cite{DoenniZolliker1996,SakaiFritsch2016}), 
transition metal dichalcogenides 
(e.g.~4Hb-TaS$_2$~\cite{RibakKanigel2019}) and 
in engineered van-der-Waals heterostructures of graphene and RuCl$_3$~\cite{MashhadiLotsch2019}. 
\begin{figure}[t]
\includegraphics[scale=1]{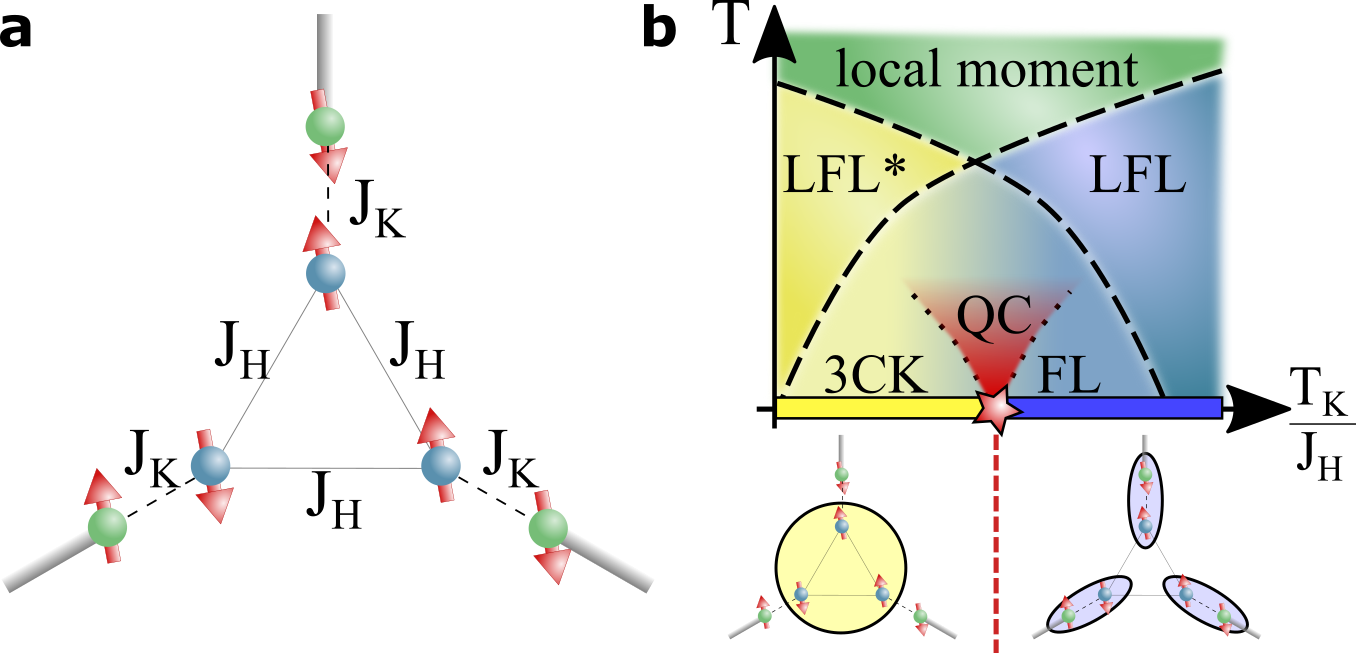} 
\caption{{\bf a} An antiferromagnetic triangle, where each spin is coupled to its own conduction bath, caricat\red{ur}es a spin-liquid competing with a Fermi liquid. {\bf b} When $T_K/J_H$ is large, each spin is individually Kondo screened (local Fermi liquid LFL, right inset). In contrast, at smallest $T_K/J_H$, 
the ground state manifold of the impurity forms an effective spin (left inset) 
and the system develops a 
3 channel Kondo (3CK) phase, in which instantons of the emergent gauge theory are irrelevant. 
Analogously to confinement in QED$_3$, instantons proliferate beyond a critical $T_K/J_H$ (red star) and restore an ordinary (L)FL.}
\label{fig:Schematics}
\end{figure}

A significant component of this intriguing physics is thought to involve 
the fractionalization of spins~\cite{Coleman1983,ArovasAuerbach1988} into
``spinons'', \yk{fractionalized particles interacting with an emergent gauge field. Fractionalization is a useful concept if
the emergent gauge theory is in the deconfining phase}. 
In (2+1)D \EJK{compact} quantum electrodynamics (QED$_3$)
deconfinement is lost via a proliferation of instantons in the gauge
field~\cite{Polyakov1977}, but the presence of fermions~\cite{HermeleWen2004,Lee2008,SongHe2019,XuMeng2019,DupuisWitczak2019} counteracts this mechanism allowing for a confinement-deconfinement quantum phase transition. It has been argued that transitions out of 
spin-liquid
phases via partial Mott delocalization, e.g.~in heavy-fermion
materials, leads to a Fermi surface reconstruction that may be
understood in these terms
~\cite{SenthilVojta2003,	Komijani18a,KomijaniColeman2018}. 

\red{\yk{In addition to} its importance for quantum materials, deconfinement of anyonic quasiparticles is of prime importance for topological quantum computation~\cite{SternLindnder2013} which may be realized in materials (as discussed above) or by artificially interweaving non-topological qubits into robust, macroscopic logical qubits, e.g. within the {surface code}. As it appears particularly desirable to \textit{electronically} manipulate and braid the emergent excitations, a natural important question regards when and how topological order is destroyed by the coupling to electronic leads.}	

Related phenomena appear in the context of magnetic
impurities, as \red{in the} overscreened 
Kondo effect
\cite{Nozieres80,Natan,TsvelikWiegmann1984,AffleckLudwig1992,ColemanTsvelik1995},
and in magnetically frustrated Kondo screened impurities
\cite{IngersentAffleck2005,HattoriTsunetsugu2012,LazarovitsSzunyogh2005,Mitchell09,ChowdhuryBulla2019}. The frustration lead to the fractionalization of the spin and an irrational
residual entropy suggestive of an underlying presence 
of non-Abelian anyons\,\cite{zinnjustin1998}. \yk{It has been recently suggested that these anyons can be potentially used as a} platform for topological quantum computation
\cite{Lopes19,Komijani19}. The common theme of these systems is an abundance of competing patterns of spin entanglement and their rearrangement at a quantum critical point (QCP). 

Here, we investigate a ``Kondo \yk{triangle} model'' involving
three antiferromagnetically coupled spins at the vertices of a triangle, each
independently coupled to its own conduction sea (Fig.~1 {\bf
a})~\cite{FerreroFabrizio2007,RamiresColeman2016}. 
The solvable limits of this model enable us to demonstrate
a transition between two distinct ground-state phases without any
symmetry breaking. \red{In one of these two phases each spin is Kondo screened separately and the spins are not mutually entangled. In the other phase, the spins are strongly entangled, the coupling to the leads results in an irrational impurity entropy.} To gain \yk{a better} insight, 
we have explored the physics of a Kondo \yk{triangle} near the large $N$ limit
of spins with an SU(N) symmetry. Our procedure contains three steps: 

\yk{i) We
pre-fractionalize spins in terms of Abrikosov fermion ``spinons''
~\cite{footnoteSpinons}.

ii) Decoupling of interactions  leads to 
a quadratic Hamiltonian \cite{AffleckMarston} with a
U(1) flux $\Phi$ through the triangle.

iii) We go beyond mean-field theory by studying {1/N corrections and} the non-perturbative effect{s} of instantons (i.e.~{phase-slips} $\Phi \rightarrow \Phi \pm 2\pi$).}

The presence of these phase-slips makes the problem distinct from the two-impurity Kondo problem discussed extensively in the past \cite{Jones88,JonesMillis1989,AffleckLudwig1992,MitchellLogan2012} \red{and allows to draw analogies to the \yk{confinement mechanism} in QED$_3$ and to fractionalization in 2+1 dimensional quantum materials in general.}

\red{We conclude this introduction with an overview of previous works on confinement-deconfinement quantum phase transitions in Kondo lattice systems. Senthil, Vojta and Sachdev\,\cite{SenthilVojta2003,SenthilSachdev2004} introduced the concept of fractionalized Fermi liquid (FL$^*$) phases, in which the Kondo screening of lattice spins 
breaks down at the expense of establishing a QSL in the spin system. When the latter is a $\mathbb Z_2$ QSL, the FL$^*$ is particularly robust, but FL$^*$ and ordinary Fermi liquid are separated by a superconducting phase which breaks particle number conservation. In contrast, the transition from an FL$^*$ with U(1) QSL to the Kondo screened Fermi liquid may be direct, does not involve the breaking of any microscopic symmetries and is governed by a quantum critical point~\cite{SenthilSachdev2004,PaulNorman2007,Vojta2010}.}
\EJK{To study this phase transition in low-dimensional Kondo problems}
\yk{
a study of fermionic degrees of freedom coupled to compact gauge fields seems essential to stabilize deconfinement. 
We are not aware of any such previous studies.}

\red{The remainder of this paper is structured as follows: In Sec.~\ref{sec:Model} we define the model under consideration and summarize the main results. Sec.~\ref{sec:3CK} contains a mapping of the \yk{triangle} model to a three channel Kondo model which is independent of the \yk{approximate} large-N treatment introduced in Sec.~\ref{sec:MF}. Fluctuation corrections beyond the $1/N$ limit are discussed in Sec.~\ref{sec:Fluct} while the conclusions, Sec.~\ref{sec:Conclusions} contain a discussion of the relationship to topological order and of the experimental implications of our work.}

\section{Model and summary of results}
\label{sec:Model}

\subsection{Model Hamiltonian}

The 
 Kondo \yk{triangle} Hamiltonian (Fig.~\ref{fig:Schematics} {\bf
a}) $H = H_c +H_{\rm H}+ H_{\rm K}$ consists of three terms:
\begin{subequations}\label{eq:H0}
\begin{align}
H_c &= \sum_{m = 1}^3 \sum_{\v p} \tilde c^\dagger_{\alpha,m} (\v p) \epsilon(\v p) \tilde c_{\alpha,m}(\v p), \label{eq:Hleads}\\
H_{\rm H} &= \frac{J_H}{N}\sum_{m = 1}^3 \hat{S}^a_m \hat{S}^a_{m+1},  \\
H_{\rm K} &= \frac{J_K}{N}\sum_{m = 1}^3 \hat{S}^a_m c^\dagger_{\alpha,m}(0) \sigma^a_{\alpha \beta} c_{\beta,m}(0).
\end{align}
\end{subequations}
The operators 
$c^\dagger_m(\v x)$ [$\tilde{c}^\dagger_m(\v p) = \sum_x e^{- i \v
p\cdot \v x} c^\dagger_m(\v x)$] create electrons 
on lead $m$, with a dispersion $\epsilon (\v p)$. 
The $\sigma^a$ ($a = 1 \dots N^2-1$) are generators of {the
fundamental representation of} SU(N) and $\hat{S}^a_m$ are the
corresponding spin operators. Summation convention over repeated spin
indices $\alpha, \beta = 1, \dots, N$ is {implied but summations over
the lead index $m$ are written explicitly. } 
\yk{In this paper, we will mainly} use the Abrikosov fermion representation of spins $\hat{S}_{m}^a =
 f_{\alpha,m}^\dagger [\sigma^a]_{\alpha \beta} f_{m,\beta}$, with the
 constraint $f_{\alpha,m}^\dagger f_{\alpha,m} = Q$ where $Q = N q$.

\subsection{Comparison to previous works}

For a large Kondo temperature $T_K  = D e^{-1/J_K \rho} \gg J_H $ the
model yields a local Fermi liquid (LFL), see Fig.~\ref{fig:Schematics}
{\bf b}, in which each spin is magnetically screened by its own conduction band.
The situation is more intricate for small {$T_K\ll J_H$}. For SU(2) spins, \red{Ferrero \textit{et al.}~\cite{FerreroFabrizio2007} employed a combination of conformal field theory and numerical renormalization group to demonstrate that} the LFL
phase is stable at all values of the ratio $T_K/J_H$. \red{Recently, we investigated the ferromagnetic version \cite{DrouinColeman2021} of Eq.~\eqref{eq:H0}, but we are not aware \yk{of} other studies of the antiferromagnetic model defined in Eq.~\eqref{eq:H0}.}

\red{However, $C_3$ symmetric models of spin-1/2 triangles which are Kondo coupled~\cite{KudasovUzdin2002} to a single 2D or 3D electronic bath were considered by Lazarovits \textit{et al.}~\cite{LazarovitsSzunyogh2005} using a renormalization group approach. Contrary to Eq.~\eqref{eq:H0}, this setup features substantial intersite correlations $\langle c^\dagger_{\alpha,m}(\tau) c_{\alpha,m+1}(0) \rangle$ which can lead to an exotic non-Fermi-liquid fixed point. \yk{The model was studied numerically by Paul and Ingersent~\cite{PaulIngersent1996} and analytically by Ingersent \textit{et al.}~\cite{IngersentAffleck2005}. Very recently Eickhoff and Anders~\cite{EickhoffAnders2020}, have re-visited the model with the goal of developing a cluster dynamical mean-field theory.}}

\red{Finally, a vast amount of literature is devoted to asymmetric triangles, in which Kondo-coupling to the leads is site selective, and/or the Heisenberg interaction is not homogeneous, see e.g.~\onlinecite{Mitchell2009,Baruselli2013,Wojcik2020}.}

\subsection{Summary of results}
\label{sec:Summary}

In this
work, we generalize this model beyond SU(2), to the case of 
spins forming an antisymmetric representation of SU(N), 
described by vertical Young tableaux with $Q$
boxes. \red{In Sec.~\ref{sec:3CK}, w}e show that for a sequence of $(N,Q)$ our model at smallest
$T_K/J_H$ maps onto a single {composite spin, overscreened by three
conduction channels, denoted here as 3CK}. This solvable limit 
corresponds to a phase with a non-trivial ground state degeneracy, differing
from the LFL at large $T_K/J_H$. Yet, neither phase
breaks any  symmetries of the model.

Within the large-$N$ approach, the appearance  of spinons
is accompanied by an emergent $U (1)$ gauge field on the links of the
triangle, with a gauge invariant flux $\oint \vec{A}\cdot
d\vec{x}=\Phi $ that threads the triangle. 
The 3CK phase (Fig.\,1{\bf b}), is characterized by the ordering of the symmetric ring exchange operator 
\begin{equation}
\mathcal O_s \equiv d_{abc} \hat S^a_1 \hat S^b_2 \hat S^c_3 \propto \cos(\Phi), \quad d_{abc} \equiv \tr [\sigma^a \lbrace \sigma^b, \sigma^c \rbrace], \label{eq:Orderparameter}
\end{equation}
 which preserves time reversal, spin SU(N) and crystalline $C_{3v}$
 symmetries. 
 By contrast, 
in the Fermi liquid (FL),  {phase-slips} proliferate, 
confining the 
spinons to each lead, and in this sense,  the two phases are separated
by a confinement-deconfinement transition. Both (FL and 3CK) phases are
robust against deformations of the triangle
(i.e.~unequal $J_H$) which make them suitable for future experimental realizations.

{Finally, we comment on special values of $Q$ and $N$. First, the particle-hole symmetric representation $Q/N=1/2$, which is related to SU(2) spins, has mean-field solutions for which some of the links are missing and the flux is ill-defined (see Fig.~\ref{fig:Instanton} {{\bf a}}, below). {Moreover,} 
the order parameter $\mathcal O_s$ of the 3CK phase vanishes \yk{for $N=2$} since $d_{abc} = 0$ for {SU(2) spins}. These arguments explain
the 
persistence of the FL phase down to $T_K/J_H \rightarrow 0$ for the SU(2) {\yk{triangle}}\,\cite{FerreroFabrizio2007}. 

Second, {at commensurate representations $Q = N/3$, or} $Q = 2N/3$, the spins form a singlet at small $T_K/J_H$ and the competition between Heisenberg and Kondo {interactions} is analogous to the \red{SU(2)} two-impurity two-channel Kondo problem, i.e.~the two limiting phases are FLs with conduction electron phase shift of $\delta_c=0,\pi$. For these commensurate representations, instead of the 3CK phase, we have a FL$^*$, i.e.~a gapped spin-liquid which is {robust to the Kondo interaction up to a threshold coupling}, and a FL$^*$ to FL transition.

\section{Mapping to three-channel-Kondo models}
\label{sec:3CK}

We first highlight a subset of
models,  with $N =  3 Q  +1$, of which the simplest is the fundamental
representation of $SU (4)$.  In these special cases, we can show that
three channel Kondo (3CK) behavior develops at large $J_H$. To see this, we first solve $H_H$ at $J_K = 0$. \red{It is convenient to employ the previously introduced Abrikosov fermion representation of the spin, and we emphasize that here no approximations are made (for details see Appendix~\ref{app:3CK})}.}
The antiferromagnetic coupling $J_{H}$ favors the formation of a
maximally antisymmetrized combination of 3Q spinons. 
Since $3Q = N-1${,} this system is one spin\red{on} short of an
overall SU(N) singlet. \yk{Indeed if one of the spins, say $m$, had a larger representation, i.e. $Q+1$ (rather than $Q$) vertical boxes, the three spins could form a singlet, denoted by $\ket{{\rm singlet},m}$. When all spins have representation $Q$, the ground state of the Heisenberg Hamiltonian $H_H$ can be shown (Appendix \ref{sec:lowenergy}) to be}
\be
\yk{\ket{\alpha}=\frac{1}{\sqrt 3}\sum_m f\dn_{m,\alpha}\ket{{\rm singlet}, m}}\label{eq:TriangleGroundState}
\ee
\red{where $\alpha = 1 \dots N$}, \yk{and states $\{\ket{\alpha}\}$ form a basis for the} conjugate
representation of SU(N). \yk{T}he corresponding
matrix elements of spin operators in the
ground state manifold are given by $\braket{ \alpha \vert
\hat{S}^a_m \vert  \alpha'} = -\sigma^a_{ \alpha', 
\alpha}$. \red{Since the groundstate of the triangle is given by a spinon hole, it is suggestive to}
also represent the conduction band
in terms of holes $c_{\alpha, m} (\v x) \rightarrow h^\dagger_{\alpha,
m} (\v x), c^\dagger_{\alpha, m} (\v x) \rightarrow h_{\alpha, m} (\v
x)$. In the limit of large $J_H$ we thus find a Kondo coupling
\begin{eqnarray}
H_K 
&=& \frac{J_K}{N} \sum_{m =1}^3 [\hat{S}^a]^T h^\dagger_{m, \alpha}(0)[ \sigma^a]^T_{\alpha \beta} h_{m,\beta}(0), \label{eq:Heff3CK}
\end{eqnarray}
between the spin and a Fermi sea of holes. Thus at large $J_{H}$, the model
\eqref{eq:H0} is equivalent to the 3CK
problem in the conjugate representation of SU(N), which is \yk{equivalent} to the 3CK Kondo model, an exactly solvable model \yk{(See Appendix \ref{sec:lowenergy})}.
From this mapping, we know that 
the {ground state has an irrational degeneracy of}~\cite{Tsvelik1985,Affleck1995,JerezAndreiZarand1998}  
\begin{equation}
g_N = 1+ 2 \cos \left(\frac{2 \pi }{N+3}\right). \label{eq:gN}
\end{equation}
\red{N}ote that $\lim_{N \to \infty}g_N =3$.
We now develop an approximate field theoretical technique which connects the two limits of the phase diagram,~Fig.~\ref{fig:Schematics}.


\section{Large-N treatment}
\label{sec:MF}

\subsection{Hubbard-Stratonovich decoupling}

\begin{figure}
\includegraphics[scale=1]{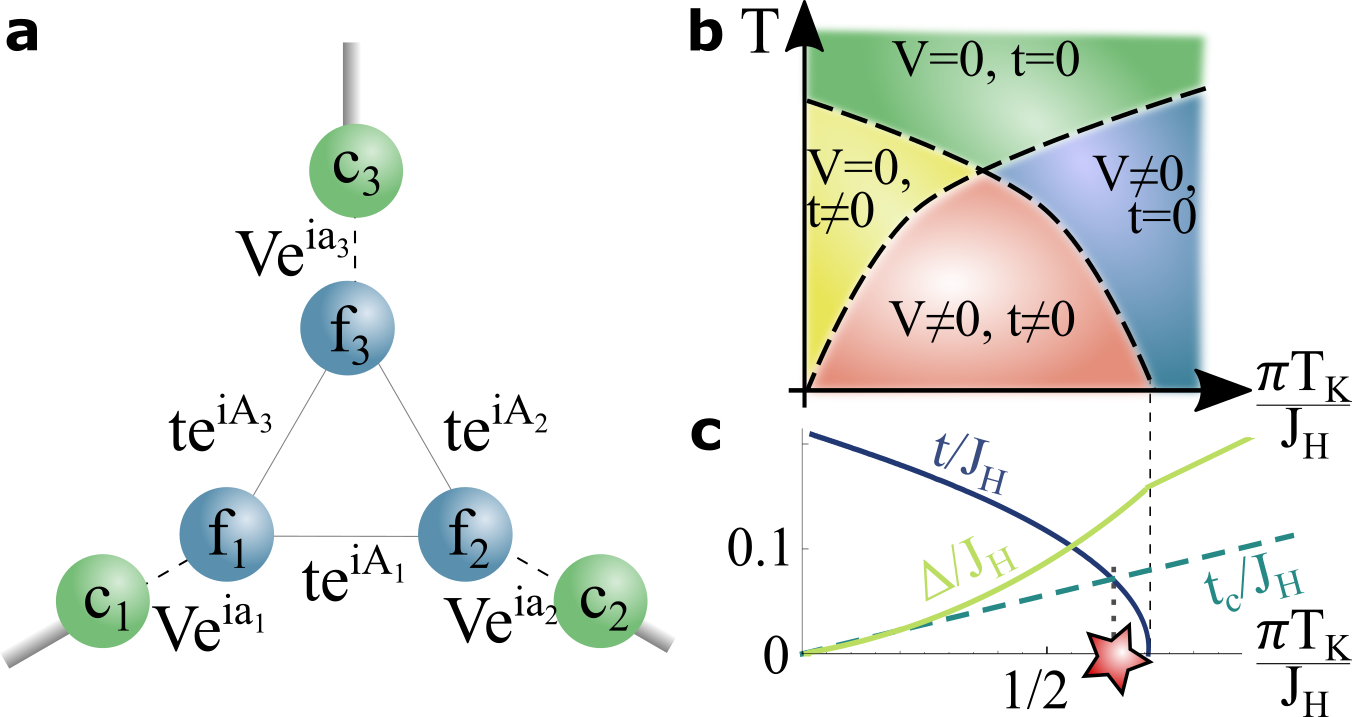} 
\caption{{\bf a} Pictorial representation of the mean-field Hamiltonian. {\bf b} \red{Schematic mean-field} phase diagram. {\bf c} 
Zero temperature, mean-field behavior of $t$, $\Delta = \pi \rho V^2$ as a function of Doniach parameter (here $J_4 = 0.3 J_H $, $J_s \approx 0.29 J_H$). The position where $t_c$\red{, defined in Eq.~\eqref{eq:tc} below}, crosses $t$ defines the confinement-deconfinement quantum phase transition at which phase slips proliferate (red star, here $N = 4$ and $q = 1/4$). }
\label{fig:Goldstone}
\end{figure}

\yk{Representing the spins using Abrikosov pseudofermions, leads to four-fermion interactions, which} 
 we decouple using Hubbard-Stratonovich transformation in the leading channels, selected by the large-$N$ limit
(see Fig.~\ref{fig:Goldstone} {\bf a} \red{for illustration})

\begin{align}
S&=S_c+\int d \tau \sum_m\Big \lbrace f^\dagger_{\alpha,m} [\partial_\tau + \lambda_m] f_{\alpha,m} -\lambda_m q N + \frac{N \vert V_m \vert^2}{J_K}\notag \\
&\hspace{-.1cm}+ \frac{N \vert t_m \vert^2}{J_H}  + [V_m f^\dagger_{\alpha,m} c_{\alpha, m} - t_m f^\dagger_{\alpha,m} f_{\alpha, m+1} + H.c.]  \Big \rbrace. \label{eq:AbrikosovAction}
\end{align}
{Here, $t_m=\abs{t_m}e^{iA_m}$ and $V_m=\abs{V_m}e^{ia_m}$} {and the Lagrange multipliers $\lambda_m$ enforce the constraint} \red{(for details, see Appendix~\ref{app:ImpurityPartition}).}

\subsection{Mean field solution}
\label{sec:MeanField}

\subsubsection{Mean field Ans\"{a}tze}

\begin{figure}
\includegraphics[scale=1]{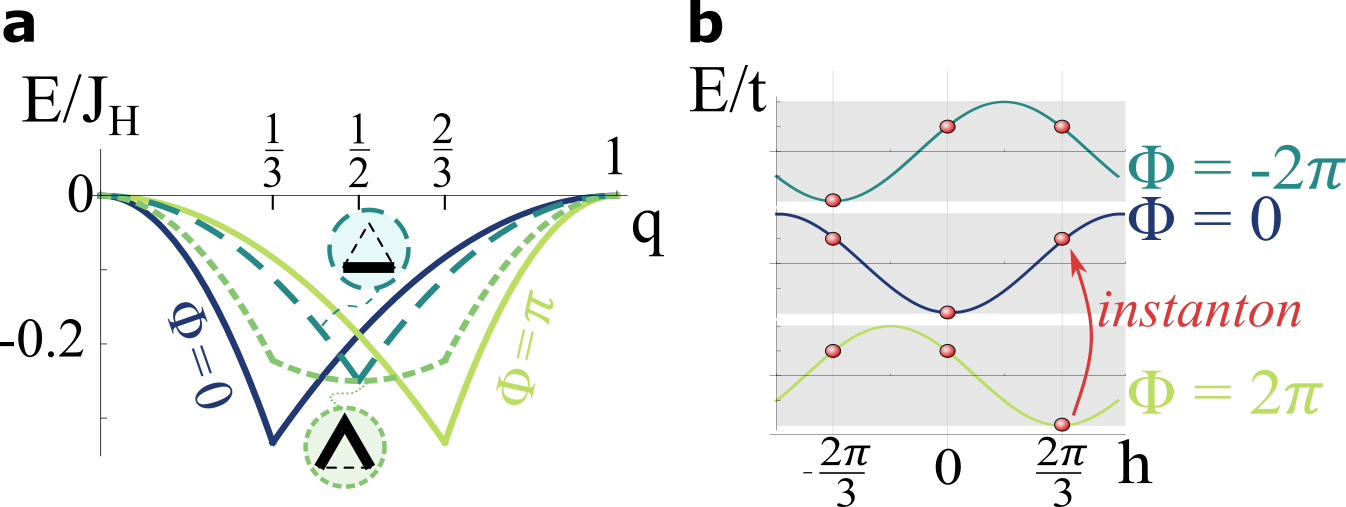} 
\caption{{\bf a} Ground state energies at $T_K = 0$ as a function of filling $q$, comparing rotationally symmetric solutions (labeled by their flux $\Phi$) with symmetry broken states with one (two) non-zero $t_m$ [graphically labeled by triangles with one (two) thick bonds]. {\bf b} Single particle spectra of spinons at $T_K = 0$ with helicity quantum numbers $h \in \lbrace 0, \pm 2\pi/3\rbrace$ for different flux configurations $\Phi = 0, \pm 2\pi$. \red{Instantons map} $\Phi \rightarrow \Phi \pm 2\pi$ \red{and thereby} reshuffle the wave functions but leave the spectrum unaltered.}
\label{fig:Instanton}
\end{figure}

In the limit $N \rightarrow \infty$, the bosonic path integral{s} can be evaluated at the saddle point level for static configurations of the fields. At $T_K = V_m = 0$, Fig.~\ref{fig:Instanton} {\bf {a}} demonstrates the stability of homogeneous solutions with $t_m = t e^{i A_m}$ and zero or $\pi$ flux $\Phi = \sum_m A_m$ away from half-filling. \red{In the reverse limit $J_H = t_m = 0$, the equality of Kondo couplings at each of the three sites implies the same hybridization $\vert V_m \vert$ for all $m$. In Read-Newns gauge the phase of $V_m$ is absorbed into $\lambda_m$, which also takes the same mean field value at each site.} 
Motivated by this, we concentrate on \red{rotationally symmetric solutions}
$V_m = V$, $\vert t_m \vert = t$ and $\lambda_m = \lambda$, {all real}, and $q<1/3$. \red{In this case, the spectrum can be \yk{found} by Fourier transformation leading to a spinon spectrum $\lambda_h = \lambda - 2 t \cos(h + \Phi/3)$, see Fig.~\ref{fig:Instanton} {\bf b}. Here, we introduced the helicity $h = 0, \pm 2\pi/3$ (the crystal momentum of the periodic 3-site chain). 
Using this solution, the fermionic path integral can be taken exactly and leads to a free energy
\begin{equation}
\frac{F}{N} = - T \sum_{\epsilon_n,h} \ln[- G_h^{-1}(\epsilon_n)] e^{i\epsilon_n \eta} + 3 \left ( \frac{t^2}{J_H} +  \frac{\Delta}{\pi \rho J_K} - \lambda q  \right ), \label{eq:FermionDeterminant}
\end{equation}
Here, $G_h^{-1}(\epsilon_n) = i \epsilon_n - \lambda_h + i \Delta \yk{sgn(\epsilon_n)}$ \yk{in the wide bandwidth limit where $\Delta = \pi \rho V^2$ is the hybridization energy related to $V$ and} the density of states $\rho$.  The variation of the free energy with respect to the parameters $\Phi$, $\lambda, \Delta, t$ leads to a set of mean field equations} \EJK{of which we discuss the solutions below.}

\subsubsection{Finite temperature phase diagram}

\red{Before presenting details about these equations at zero temperature, we discuss the finite temperature} 
%
%
{mean field phase diagram, Fig.~\ref{fig:Goldstone} {\bf b} \red{(a calculation of mean field transition temperatures is presented in App.~\ref{app:FiniteTempMeanField})}. 

(i) At the highest temperature, the spinons are
decoupled, both from each other ($t = 0$) and from their respective
conduction band ($V = 0$), so the impurity spins are neither entangled
nor screened. \yk{This is characterized by decoupled spins showing Curie susceptibility behavior}.

(ii) For $T < T_K$ and large $T_K/J_H$, all moments are
individually screened (LFL), \yk{i.e. $t=0$ but $V> 0$.} 

(iii) At smallest $T_K/J_H \ll 1$ and
finite temperature, $t >0$ but $V =0$: here a  miniature spin-liquid
behavior develops. Since the phase shift for all conduction bands is
zero, we denote this phase ``LFL$^*$''~\cite{SenthilVojta2003} in
Fig.~\ref{fig:Schematics} {\bf b}. 

(iv) Finally, the mean-field phase
n which both $V >0$ and $t >0$, \yk{which} is the focus of the rest of the paper. \yk{We will show that there is a} \EJK{deconfinement transition} 
\yk{inside this} \EJK{mean field} \yk{phase.} 

\yk{Next, we derive the mean-field transition between these} \EJK{zero temperature} \yk{phases to map out the mean-field phase diagram}.
\subsubsection{Zero temperature mean field equations}

\red{We now \yk{investigate} the two zero temperature phases. We readily find that $\Phi = 0$ solves the mean field equations and we concentrate on this solution for $q <1/2$. It is convenient to replace the three other variational parameters $(\lambda, \Delta, t)$ by $(\delta_0, \delta_{2\pi/3}, t)$, where $\delta_h = \delta_{-h} = \text{arccot}(\lambda_h/\Delta)$ is the phase shift in the helicity channel $h$. The variation of the free energy with respect to the Lagrange multiplier $\lambda$ enforces a sum rule 
}\red{
\begin{subequations}\label{eq:MFCondAll}
\begin{equation}
3\pi q = \delta_0 + 2 \delta_{2\pi/3},
\end{equation}
while the variation with respect to $t$ connects the difference $d = \delta_{2\pi/3} - \delta_0$ of phase shifts with the spinon hopping
\begin{equation}
3 \pi t = -J_H d. \label{eq:t}
\end{equation}
\EJK{Note that $t>0$ implies $d <0$. }
The third saddle point equation follows from the variation of the action with respect to $\Delta$. We exploit the previous equation and obtain
\begin{equation}
 \left( \frac{\sin(d)}{d} \frac{\pi T_K}{J_H}\right)^3 = \sin \left (\frac{3 \pi q +d}{3} \right)\sin^2 \left (\frac{3 \pi q-2d}{3} \right). \label{eq:MFCond}
\end{equation} 
\end{subequations}
}\red{
Note that there is only one variational parameter, $d$, in this equation, while $q$ and $\pi T_K/J_H$ are fixed externally. For a graphical solution of Eq.~\eqref{eq:MFCond}, see Fig.~\ref{fig:MeanFieldSol}. It demonstrates that for $q = 1/3$, a state where both $t \neq 0$ and $V \neq 0$ is never the ground state, while for $q < 1/3$ of prime interest in this work,  there is a phase with $t \neq 0$ and $V \neq 0$ which
}
persists to the smallest $T_K/J_H$ and is separated
from the LFL by a first-order phase transition (an artifact of the \red{mean-field} approach). 

\begin{figure*}
\includegraphics[scale=1]{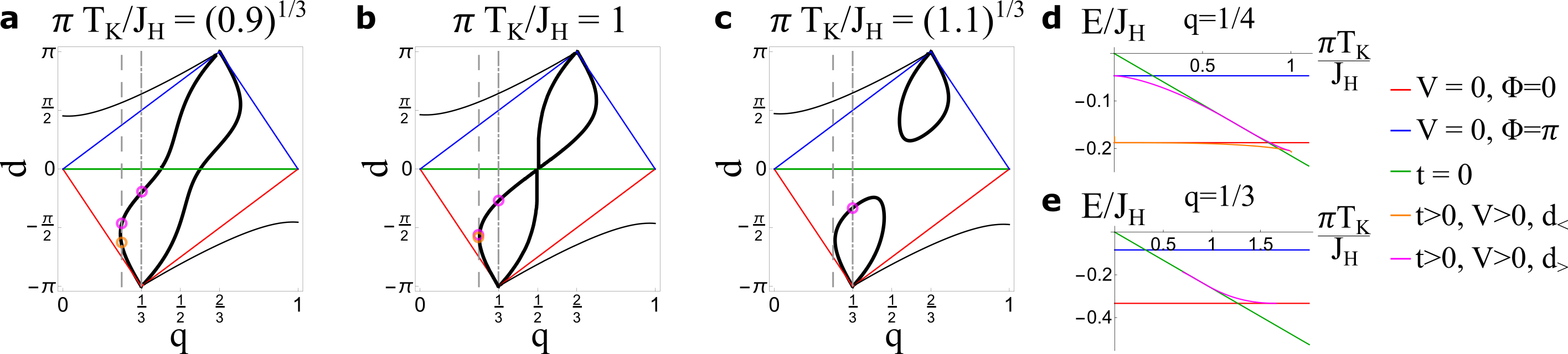}
\caption{\red{Graphical illustration of mean field solutions. {\bf a} - {\bf c} Each point on the black curves corresponds to a solution of Eq.~\eqref{eq:MFCond} for a given value of the Doniach parameter $\pi T_K/J_H$. \EJK{For $q <1/2$ ($q >1/2$), we concentrate on $d<0$ ($d>0$) (other solutions are ``false vacua'' indicated as thin lines).}
For comparison, we include analogous curves of solutions in the limit $T_K = 0$  [$t>0, V = 0$ (red) and $t<0, V = 0$ (blue)] as well as $J_H = 0$ [$t = 0, V>0$ (green)].
The vertical gray dashed lines indicate the position of $q = 1/4, 1/3$. For $q = 1/4$ there are up to two non-trivial solutions with $- \pi <d <0$ (orange and pink circles, denoted $d_<$ and $d_>$, respectively), while for $q =1/3$ there is only one non-trivial solution (pink circle) in addition to the solution $d = - \pi$ corresponding to complete Kondo breakdown ($V =0$). The corresponding mean field energy is plotted with the same color code in {\bf d, e}, and compared to the solutions where either $V = 0$ or $t = 0$. }} 
\label{fig:MeanFieldSol}
\end{figure*}


\subsection{Symmetries}
\label{sec:Symmetries}

\red{Here, w}e summarize the underlying symmetry breaking \red{using the language of conventional phase transitions. We emphasize, however, that no physical symmetry is broken in either 3CK or LFL phase. A parallel discussion in terms the projective symmetry groups is therefore included in Sec.~\ref{sec:GaugeAndProj}, below. }
   
In the
UV {($t=V=0$)}, Eq.~\eqref{eq:AbrikosovAction} displays a symmetry
$U_f(1)^{\otimes 3} \times U_c(1)^{\otimes 3}$ (i.e.~$f_m \rightarrow
e^{i \phi_m(\tau)} f_m$, $c_m \rightarrow e^{i \varphi_m} c_m$), of
which $U_f(1)^{\otimes 3}$ is a gauge symmetry. The Kondo effect on
each site ${m}$ {($V_m\neq 0$)} breaks the symmetry as
$U_f(1)^{\otimes 3} \times U_c(1)^{\otimes 3} \rightarrow U_{\rm
cf}(1)^{\otimes 3} {\equiv G}$ fixing $\varphi_m = \phi_m$ for each
spin/bath $m$ separately. {The three {``Goldstone modes''} are eaten
by the {Lagrange multiplier} $\lambda_{m}$ within the {Read-Newns
gauge} \cite{ColemanMarston}.} Most interesting for the present study
is the establishment of a spin liquid, in which $\vert t \vert>0$
fixes $\phi_m = \phi + j(m-1) \frac{2\pi}{3}$ with $j \in \{0,1,2\} =
\mathbb Z_3$.  {Thus, the remaining symmetry $H = U_{\rm cf}(1) \times
\mathbb Z_3$ is generated by the total phase $\phi$ and the insertion
of} a total flux of $2\pi j$, i.e.~a {large} gauge transformation
which leaves the spectrum unchanged, but re-arranges the eigenstates{,
Fig.~\ref{fig:Instanton} {\bf b}.  The symmetry breaking $G
\rightarrow H$} is {apparent within the} Landau free energy, which we
derived (see Appendix~\ref{app:Landau}) at $T = 0$, $V>0$ and small $\bar t = {\vert
t_m \vert}/T_K$
\begin{equation}
\frac{F}{N} =  T_K  \left [ \alpha \bar t\hphantom{.}^2 - \beta \bar t\hphantom{.}^3 \cos(\Phi) + \gamma \bar t\hphantom{.}^4 {+O(\bar t^5)}\right ]. \label{eq:Landau}
\end{equation} 
where {$\alpha = 3 T_K/J_H - \sin(\pi q)/\pi$}. {The flux $\Phi = \sum_{m} A_m \in [0,6\pi)$, but Eq.~\eqref{eq:Landau} is $2\pi$ periodic in the total flux, pointing to the emergent $\mathbb Z_3$ gauge symmetry of the problem.} 

\subsection{Bilinear coupling and ring exchange}

\red{The Landau free energy illustrates the first order nature of the mean-field transition: The cubic term is a consequence of the threefold symmetry of the impurity, and additionally the microscopic parameters in Eq.~\eqref{eq:H0} imply $\gamma <0$ near the transition, reinforcing the first order behavior. A negative quartic term is typical in the large-N treatments and can be cured by inclusion of a biquadratic interaction \cite{Komijani18a}
\begin{equation}
H_4 = -\frac{\pi^3J_4}{2N^3} \sum_m [\hat{S}_m^a \hat{S}_{m+1}^a]^2,
\end{equation}
leading to $\gamma = 3[J_4 \sin(\pi q)^4/T_K - \sin(3\pi q)]/(2\pi)$.
The first order jump is further weakened by the addition of a totally symmetric ring exchange 
\begin{equation}
H_3 = -\pi^2\frac{J_s}{N} d_{abc} \hat{S}_1^a\hat{S}_2^b\hat{S}_3^c,
\end{equation}
so that microscopically $\beta =  [\sin(2\pi q) - J_s \sin(\pi q)^3/T_K]/\pi$ after integration of fermions. A similar integration is the origin of the \yk{relation} $\mathcal O_s \sim \cos(\Phi)$ presented in Eq.~\eqref{eq:Orderparameter}.}

Ring exchange terms \yk{can be} employed to physically access the emergent gauge flux $\Phi$. An adiabatic flux insertion \yk{can be achieved by adiabatically tuning $\theta(t)$} in 
\be
\yk{\Delta} H\yk{(t)}= {-}\frac{J}{N}\Big\{d_{abc} \cos\theta\yk{(t)}+  f_{abc}\sin\theta\yk{(t)}\Big\} \hat S_1^a \hat S_2^b \hat S_3^c \notag
\ee
\yk{where} {$d_{abc}$ and $f_{abc}$ are symmetric and antisymmetric} structure factors of SU(N). \yk{A mean-field decoupling of $H+\Delta H$ leads to $\Delta F {\propto - J} \cos(\Phi - \theta)$}.
\subsection{A study of projective symmetry group}
\label{sec:GaugeAndProj}

\red{\yk{We now return to} the emergent gauge invariance in the problem and employ the method of projective symmetry groups (PSGs), introduced \cite{Wen2002} to categorize gapless spin liquid states which do not break any microscopic (e.g. crystalline) symmetries. \yk{To recapitulate the procedure:}} 

\red{(i) Consider a mean field tight binding model of spinons $f_{\alpha,m}$, in our case Eq.~\eqref{eq:AbrikosovAction}. Because spinons carry an emergent gauge charge, mean field tight-binding models which can be transformed into each other by means of a gauge transformation are equivalent.}

\red{(ii) The group of microscopic symmetry operations \yk{followed by a gauge transformation} which leave the tight binding model invariant form the PSG of the model.} 

\red{(iii) The subgroup of gauge transformations which leave the tight binding Hamiltonian invariant form the invariant gauge group (IGG).} 

\red{(iv) The actual symmetry group (SG) of the model is thus SG = PSG/IGG. Hence the PSG can be seen as an extension of the SG.} 

\red{The IGG also} \EJK{places constraints on}
\red{Wilson loop operators $P_C$~\cite{Zhou17}, which are products of Peierls gauge \EJK{fields} along closed contours on links of the lattice ($P_C =\prod_{m = 1}^3 e^{i A_m}$ in our simple three-site lattice). 
Wilson loops \EJK{are} a particularly useful definition for gapless topological quantum states when standard signatures (such as a degenerate ground state manifold) are less obvious.}

\red{As mentioned, in our case the} \EJK{infrared gauge transformations are $f_{\alpha, m} \rightarrow e^{i \phi_m} f_{\alpha, m}$, $c_{\alpha, m} \rightarrow e^{i \phi_m} c_{\alpha, m}$, $A_m \rightarrow A_m + \phi_m - \phi_{m + 1}$ and imply an IGG which is U(1).} 
\red{The crystalline symmetries group of the triangle 
is generated by 120$^\circ$ rotations $R$, $R^3 = 1$ and an involutory mirror operation $M = M^{-1}$ exchanging sites $m = 1 \leftrightarrow m = 2$. They do not commute, instead $MRMR = 1$. We may then proceed with the analysis of the PSG assuming deconfining gauge fields. To projectively represent the rotation, we perform a gauge transformation $f_m \rightarrow G_R(m) f_m$, $c_m \rightarrow G_R(m) c_m$ with $G_R(m) = e^{i A_m}$ after application of the crystalline symmetry operation. Analogously, the mirror exchanging sites $1 \leftrightarrow 2$ is projectively represented by employing $G_M(m) = e^{i (A_3 - A_2)\delta_{m,3}}$. Since we assume time reversal symmetry, there is a gauge in which all hopping matrix elements are real, i.e.~$A_m~\in~\{ 0, \pi \}$. Then the algebra of projective symmetry operations is $(G_R R)^3 = e^{i\Phi} =\pm 1$, $(G_M M)^2 = 1$, $(G_M M)(G_R R)(G_M M)(G_R R) = 1$. Thus, the two mean field states associated to $\Phi = 0, \pi$ in Fig.~\ref{fig:Instanton} {\bf a} are categorized by different algebraic PSGs. On the mean field level these two states are separated by multiple symmetry broken states - this is reminiscent of the transition between 2D quantum phases with different PSGs coupled to fermionic matter~\cite{KoenigTsvelik2019}. }

\section{Fluctuations and gauge fields}
\label{sec:Fluct}

\red{In the previous section we discussed the mean field solution to Eq.~\eqref{eq:H0}, which is valid at $N = \infty$. Here\yk{,} we consider fluctuation corrections beyond this limit. }

\subsection{Dynamics of low-energy excitations}

\red{The bosonic low-energy excitations in the model are the phases $A_m$, whose action may be derived microscopically by \yk{a lengthy but} straightforward integration of fermionic degrees of freedom, see Appendix~\ref{app:Fluctuations}, leading to $S[A_m] = S_{\rm diss} + S_{\rm Maxwell}$ 
\begin{subequations}
\begin{align}
S_{\rm diss} &= \int \frac{d \omega}{2\pi} \frac{\eta}{4\pi} \Phi(\omega) \Phi(-\omega) \vert \omega \vert, \label{eq:Sdiss} \\ 
S_{\rm Maxwell} &= \int d\tau \sum_m \frac{\epsilon}{2}\dot{A}_m^2, \label{eq:GaugeAction}
\end{align}
\label{eq:SA}
\end{subequations}
where 
\begin{subequations}
\begin{align}
\eta &\simeq 3 N   \frac{t^2}{T_K^2} \sin^2(\pi q), \quad (\text{for} \; t \ll T_K \cos(\pi q)), \label{eq:eta}\\
\epsilon &= \frac{2N}{9J_H} \left (1 + \frac{J_H\sin(\delta_{2\pi/3})^2}{2\pi \Delta}\right ) \left (1 + 2 \rho \Delta \right ).  \label{eq:eps}
\end{align}
\end{subequations}
Here, we presented the microscopic expression for the low-energy ($\vert \omega \vert \ll T_K$) dissipative dynamics of $\Phi$ in the limit $t \ll T_K$ (near the QCP, see Fig.~\ref{fig:Goldstone} {\bf c}). For a more comprehensive expression, see Eq.~\eqref{eq:SPhi} of the Appendix.}

Before \yk{discussing} the features of this emergent gauge theory, we analyze the fluctuations using the more conventional language of Goldstone bosons.
}

\subsection{Goldstone bosons}

\red{When $t/t_K >0$, t}wo combinations of $A$ phases, parametrized by $A_m =  - 2 \vec x \cdot \hat e_m/3$, where {$\hat e_{1,2} = (\pm \sqrt{3},1)/2$, $\hat e_3 = (0,-1)$, {$\vec x= (x_1,x_2)$}, are zero modes of \red{the free energy} Eq.~\eqref{eq:Landau} \red{and parametrize the manifold of Goldstone bosons (see Sec.~\ref{sec:Symmetries})
\begin{equation}
\frac{G}{H} = \frac{U(1) \times U(1)\times U(1)}{U(1) \times \mathbb Z_3}. \label{eq:GoldstoneMF}
\end{equation}
In contrast, the third linear combination of phases, $\Phi = \sum_m A_m$, is gapped. For $J_H \gg T_K$, the two brackets entering $\epsilon$ in Eq.~\eqref{eq:eps} are approximately one, so we omit them for simplicity. The} effective action \red{of Goldstone bosons is thus}
\begin{equation}
S_{\rm Goldstone}[\vec x] = \int d \tau  \frac{m_x \dot {\vec x}^2}{2}, \label{eq:GoldstoneAction}
\end{equation}
 where $m_x = 8 N/(27 J_H)$. \red{This action describes a free particle with position $\vec x$ and mass $m_x$ living on the flat, yet compact manifold \eqref{eq:GoldstoneMF}.} The ground state is an $\vec x$-independent wave function and, due to the compactness of \red{$G/H$}, detached from the first excited state at energy $\red{\sim}1/m_x$.
  
\red{As a consequence, d}espite the mean field value $t>0$, intersite Green's functions $\langle c^\dagger_{m} c_{m+1} \rangle \sim \yk{\langle  e^{i A_m} \rangle}$ vanish upon integration of Goldstone modes. \red{Therefore, the absence of charge transfer between different leads is ensured by the fluctuations beyond the $N \rightarrow \infty$ limit. Equivalently, \yk{this} can be interpreted as} a consequence of gauge symmetry \EJK{which impedes charge fluctuations on the impurity sites}. 

\subsection{Confinement-Deconfinement transition}
\label{sec:ConfDeconfTrans}

So far, we 
incorporated leading terms in a $1/N$ series. Now, we address processes with Boltzmann weight $\Gamma \sim e^{-N}$ (instantons). Naively, these are strongly suppressed, yet we demonstrate a proliferation of instantons at sufficiently large $T_K/J_H$. Instantons in gauge theories are non-trivial gauge field configurations which are bound to be a pure gauge at infinity. In the present case, these 
are phase slips, i.e.~configurations of the field $\Phi(\tau)$ such that $\Phi(\infty) - \Phi(-\infty) = \pm 2\pi$, and we estimate their bare tunneling action 
{$\Gamma \sim e^{- N\bar t}$ for $\beta \ll \bar t$} \red{in Appendix~\ref{app:Estimate}}.

Considering Eq.~\eqref{eq:AbrikosovAction} with static fields $t, V$, we {can} artificially introduce~\cite{ColemanTsvelik1995} an additional Hilbert space associated to $\Phi = 0, 2\pi, 4\pi$. \yk{To manifestly illustrate the effect of phase slips, we define $\omega = e^{i 2\pi/3}$ and the following two matrices in the space of groundstate manifold
\be
\sigma_\Phi=\matc{ccc}{1 \\ & \omega \\ && \omega^2}, \qquad \tau_\Phi=\matc{ccc}{0 & 0 & 1 \\ 1 & 0 & 0 \\ 0 & 1 & 0}.
\ee
Here, $\tau_\Phi$ are clock matrices $\sigma_\Phi \tau_\Phi= \omega \tau_\Phi \sigma_\Phi $, $\tau_\Phi^3 = 1$. The phase slips accompanying spinon hoppings, are taken into account by replacement $t \rightarrow t \sigma_\Phi$.}

The infinite resummation of phase slips \red{of the latter in the partition sum leads to an} 
effective Hamiltonian derived in Appendix~\ref{app:InfResumSlips}
\begin{align}
H_{\rm eff}&= [H_c + \sum_m \lambda (f^\dagger_{\alpha, m}f_{\alpha, m} - Q) ]\mathbf 1_\Phi- \Gamma (\tau_\Phi+\tau_\Phi^{-1})  \notag\\
&+ \sum_m \left[V f^\dagger_{\alpha,m} c_{\alpha, m} \mathbf 1_\Phi -t f^\dagger_{\alpha,m} f_{\alpha, m+1}\sigma_\Phi  + H.c.\right ]. \label{eq:Heff}
\end{align}
In the formulation of Eq.~\eqref{eq:Heff}, two limiting cases become apparent. {First,} $\Gamma{/t} \rightarrow 0$ {representing} the 3CK phase. {Second, perturbation about} $\Gamma{/t} \rightarrow \infty$ {demonstrates that $t$ is RG irrelevant and the LFL is restored.}

To study the transition between these two limiting phases, we consider the {helical} (i.e.~Fourier transformed) basis $\tilde f_{\alpha,h} = \sum_m e^{-i h m} f_{\alpha,m}/\sqrt{3}$. A phase slip $t \rightarrow \omega t$ is equivalent to the instantaneous jump of the spinon energy $(t,-2t,t) \rightarrow (t,t,-2t)$ for $h = (-2\pi/3,0, 2\pi/3)${, \red{Fig.~\ref{fig:Instanton} {\bf b}}. Due to hybridization with conduction electrons a phase slip triggers an Anderson {orthogonality} catastrophe and thereby 
logarithmic attraction {of opposite phase slips, $\Delta \tau$ apart,} with an effective action 
\begin{equation}
S_{\rm slips} = \kappa \ln \vert\Delta \tau\vert.
\end{equation} 
Here
\begin{equation}\label{eq:Kappa}
{\kappa} = \frac{2N}{\pi^2} \left [ \arctan \left ( \frac{3t \Delta }{\Delta^2 + \lambda^2} \right) \right]^2
\end{equation} 
\red{is the stiffness of interaction as determined by} the perturbative inclusion of a single \yk{pair} 
of \yk{opposite} phase slips \red{at distance $\Delta \tau$ in the fermionic partition sum, see Appendix~\ref{app:InstIA}. Integration over $\Delta \tau$} leads to the free energy 
\begin{equation}
F = \ln(g_N) T + \mathcal C ({\Gamma^2}/{\lambda}) (T/\lambda)^{\kappa - 1},
\end{equation}
where \red{we included the effect of} 
the ground state degeneracy \red{$g_N$, Eq.~\eqref{eq:gN},} and ${\cal C}$ is a constant. This signals a quantum phase transition when the phase-slips overpower the first term at $\kappa = 2$, corresponding to, Fig.~\ref{fig:Goldstone} \red{\bf c},
\begin{equation}
t_c \sim T_K \sin (\pi q)/\sqrt{N}. \label{eq:tc}
\end{equation}
{The residual entropy at the QCP is enhanced to $S=\ln(g_N)+{\cal C}\Gamma^2/\lambda^2+{\cal O}(\Gamma^4/\lambda^4)$ {by the instanton contribution,} in consistency with the g-theorem \cite{gtheorem}.}
{T}he present model of logarithmically interacting particles on a ring of circumference $1/T$ can be cast into renormalization group language~\cite{NarayanShastry1999}: $\Gamma$ renormalizes to infinity (zero) for $t<t_c$ ($t_c<t$). However, contrary to the Berezinskii-Kosterlitz-Thouless transition, the stiffness $\kappa$ does not flow. 

So far, the deconfinement transition was studied by 
first locking $\Phi$ into one of the minima of Eq.~\eqref{eq:Landau} and subsequent
perturbative inclusion of phase slips. 
The same transition may also be {studied} in a dual language (approaching the red star of Fig.~\ref{fig:Schematics} {\bf b} from the right). In this case $\Phi$ is free to fluctuate and $\beta \ll 1$ is considered as a perturbation. From this perspective, the 3CK (FL) is the phase where $\beta$ is relevant (irrelevant). Crucially, near the transition, the dynamics of the $\Phi$ field is overdamped due to the interaction with the conduction bath, see Eq.~\eqref{eq:Sdiss}, \eqref{eq:eta}
The problem of dissipative tunneling, i.e.~$S = S_{\rm diss} - \int d \tau N T_K \beta \cos [\Phi(\tau)]$, \yk{for small $\beta$ yields a scaling equation 
\begin{equation}
\frac{d \beta}{ d\ell} = \left (1-\frac{1}{\eta} \right) \beta
\end{equation}
where $d\ell=-\log D$ in terms of the running cut-off}~\cite{CaldeiraLeggett1983,KaneFisher1992}, while the non-analytical nature of the ``kinetic'' (i.e.~damping) term is believed to prevent a renormalization of $\eta$ to all orders~\cite{NarayanShastry1999}. The condition $\eta > 1$ for relevant $\beta$ is parametrically equivalent to $t > t_c$, with $t_c$ given \red{in Eq.~\eqref{eq:tc}}. In the dual language {it is manifest that} Goldstone bosons {$\vec x$} do not affect the nature or position of the transition {because} they are by construction perpendicular to $\Phi$. 

\subsection{\yk{3CK} phase in fractionalized language}
 Before \red{concluding}, we briefly reiterate the connection to the three channel Kondo problem for $T_K \ll J_H$ {in fractionalization language}: 
 
 In this limit it is convenient to evaluate Eq.~\eqref{eq:Heff} in a gauge in which $t$ is real and positive.
Since $\Gamma$ is irrelevant in this phase, $\sigma_\Phi$ is conserved. We project on the ground state {(zero helicity $h = 0$)} of the $f$-electrons (Fig.~\ref{fig:Instanton} {\bf a}) and obtain the effective Kondo Lagrangian
\be
\yk{{\cal L}_{\rm Kondo}=\sum_m\Big[(\tilde f_{\alpha,0}^\dagger V_m c_{\alpha,m} + c.c.)+{\frac{\abs{V_m}^2}{J_K}\Big]},}
\ee
with constraint $\tilde f_{\alpha,0}^\dagger \tilde f_{\alpha,0} = 3Q$. As anticipated previously, three channels of conduction electrons are screening a single spin and 3CK physics is expected. The soft modes associated with rotations of $\vert V_m \vert$ are gapped for $T_K/J_H>0$ enforcing $V_m =Ve^{ia_m}$. \yk{Based on this observation, we conjecture that the physics discussed here for the 3CK phases of our Kondo triangle applies more generally to single-impurity three-channel Kondo systems and more generic oversceened Kondo problems.}


\section{Conclusion}
\label{sec:Conclusions}

We conclude with a discussion of multi-channel Kondo \yk{phases} as representatives of topological order and of possible experimental and numerical implications of our findings.} 

\subsection{Signatures of topological order}

While there is no magnetic ordering in any of the phases, the symmetric ring exchange operator $\mathcal O_s = d_{abc} \hat S_1^a \hat S_2^b \hat S_3^c$
displays order \red{in the 3CK phase. The ordering of such a composite operator is}
similar to \textit{order by disorder}~\cite{Villain1977,FradkinSusskind1978,ChandraLarkin1990} or \textit{vestigial order}~\cite{FernandesSchmalian2019} phenomena \red{and would suggest a characterization of the 3CK phase in terms of a generalization of spontaneous symmetry breaking.}

\red{However, we here propose a different interpretation and put forward the hypothesis that multi-channel Kondo states display a form of topological order which is similar to the quantum order in gapless QSLs. For the 3CK phase scrutinized here, the evidence is as follows:}

\red{First, as mentioned} the 3CK phase does not break any of the physical symmetries in the original model~\eqref{eq:H0}, \red{even when the $\mathcal O_s$ orders. \red{This invalidates any interpretation of the 3CK in terms of spontaneous symmetry breaking - instead we have presented a categorization using the projective symmetry group.} Second, r}egarded as an operator in the gauge theory, $\mathcal O_s {\sim \text{Re}(e^{i \sum_m A_m}})$ is a miniature Wilson loop. \red{In macroscopically extended systems, this would be taken as a clear signal of deconfinement. Third, the order of $\mathcal O_s$ is destroyed by the proliferation, or ``condensation'', of monopoles in the FL, which on the other hand are gapped in the 3CK phase. This is reminiscent of the situation in QED$_3$, while the expulsion of topological defects is generically a defining characteristic of topological states.\cite{Sachdev2018} Finally, the 3CK displays an} irrational ground state degeneracy indicating \red{gapless} anyonic excitations\red{, another striking signature of topological order.}

\red{At the same time, multichannel Kondo states are often unstable towards anisotropic coupling to the leads (see, e.g. Refs.~\onlinecite{BeriCooper2012,KoenigTsvelik2020} for exceptions). As mentioned, while the 3CK phase studied here is stable for unequal $J_H$ it is unstable if $J_K$ are unequal. This} suggests the interpretation of the 3CK phase as a symmetry protected topological state of matter \red{or as a deconfined quantum critical \yk{fixed point}}.

\subsection{Relevance for experiment and numerics}

Beyond its purpose as a\red{n analytically tractable} toy model, our investigations are relevant to the simplest cluster-dynamical mean field theory \cite{DMFT,Maier2005} approaches to Hubbard models on triangular lattices \red{which have enjoyed increased interest in recent times~\cite{KeselmanJian2020,WietekGeorges2021}. The SU(4) case studied here might be of importance for} 
twisted bilayer graphene~\cite{XuBalents2018,HauleHaule2019} with approximate valley symmetry. \red{Emergent SU(4) symmetric spin interactions~\cite{PatiKhomskii1998,LiZhang1998,TokuraNagaosa2000} were also recently predicted in spin-orbit coupled transition metal} \EJK{trihalides} \red{with low-lying $J_{\rm eff} = 3/2$ quartets \cite{YamadaJackeli2018}.} \red{SU(N) symmetric interactions of strongly correlated fermions with large flavor number $N$ have moreover been realized in cold atomic quantum emulators\,\cite{GorshkovDelRey2010,ScazzaFolling2014}.}
We conclude with the prospect of directly probing the presented theory in quantum dot experiments: Recent advance on SU(4) impurities~\cite{KellerGoldhaberGordon2014}, \yk{triangle}~\cite{SeoMahalu2013}, and three channel~\cite{IftikharPierre2018} Kondo physics may allow to artificially fabricate the setup Fig.~\ref{fig:Schematics} {\bf a} and thereby conduct an experimental study of the deconfinement transition.

\acknowledgments
 The authors appreciate discussions with
P. Chandra, \EJK{W. Metzner, Th. Sch\"afer} and E. Sela. E.J.K. was  supported by DOE Basic Energy Sciences grant DE-FG02-
99ER45790. P.\,C.\,and Y.\,K.\, were support by } NSF grant DMR-1830707. 

\appendix
\section{Mapping to three channel Kondo problem}
\label{app:3CK}

This section is devoted to the mapping of the frustrated triangle to a three channel Kondo problem \red{and contains details for Sec.~\ref{sec:3CK} of the main text}. This mapping is possible for the sequence of models with $SU(4), SU(7), SU(10) \dots$ (i.e.~$N \in 3 \mathbb N +1$) symmetry at filling $q = 1/4,2/7,3/10, \dots$ (i.e.~$Q = (N-1)/3$) and is valid when $J_H$ is the largest scale.

\subsection{Solution of the triangle alone}

We represent a given spin configuration with fixed particle number per site $Q = (N-1)/3$ by
\begin{align}
&\ket{\alpha_1\dots \alpha_{Q};\alpha_{Q+1}\dots \alpha_{2Q}; \alpha_{2Q+1}\dots \alpha_{3Q} } \notag \\
&= f^\dagger_{1,\alpha_1}\dots f^\dagger_{1,\alpha_{Q}} f^\dagger_{2,\alpha_{Q+1}} \dots f^\dagger_{ 2,\alpha_{2Q}}f^\dagger_{3, \alpha_{2Q+1}} \dots f^\dagger_{3, \alpha_{3Q}} \ket{\underline 0}. \label{eq:largeNstate}
\end{align} 
In this manifold, the spin is faithfully represented as
\begin{eqnarray}
\hat{S}^a_m = f^\dagger_{\alpha,m} \sigma^a_{\alpha \beta} f_{m,\beta}.
\end{eqnarray}
We next act on Eq.~\eqref{eq:largeNstate} with the Hamiltonian
\begin{align}
H_H &= \frac{J_H}{N}\sum_{m = 1}^3 \Big (f^\dagger_{m, \alpha}  f_{m, \beta} f^\dagger_{m+1, \beta} f_{m+1, \alpha} \notag \\
&- \frac{f^\dagger_{m, \alpha}  f_{m, \alpha} f^\dagger_{m+1, \beta} f_{m+1, \beta}}{N} \Big).
\end{align}
The last term yields a mere shift of energy $3 Q^2 J_H/N^2$ for any of the states Eq.~\eqref{eq:largeNstate}, so we omit it.
The action of the first term is the sum of permutations of two spin indices from adjacent sites, 

\begin{widetext}
\begin{align}
H_H &\ket{\alpha_1\dots \alpha_{Q};\alpha_{Q+1}\dots \alpha_{2Q}; \alpha_{2Q+1}\dots \alpha_{(N-1)} }= \frac{J_H}{N} \Big \lbrace \ket{\alpha_{Q+1}\dots \alpha_{Q};\alpha_1 \dots \alpha_{2Q}; \alpha_{2Q+1}\dots \alpha_{(N-1)} } \notag \\
&+ \ket{\alpha_1, \alpha_{Q+1} \dots \alpha_{Q};\alpha_2, \alpha_{Q+2}\dots \alpha_{2Q}; \alpha_{2Q+1}\dots \alpha_{(N-1)} } + \text{(similar perm. between sites 1,2)} \notag \\
&+\ket{\alpha_1\dots \alpha_{Q};\alpha_{2Q+1}\dots \alpha_{2Q}; \alpha_{Q+1}\dots \alpha_{(N-1)} } + \text{(similar perm. between sites 2,3)} \notag \\
&+\ket{\alpha_{2Q+1} \dots \alpha_{Q};\alpha_{Q+1}\dots \alpha_{2Q}; \alpha_1\dots \alpha_{(N-1)} } + \text{(similar perm. between sites 3,1)} \Big \rbrace.
\end{align}

Therefore, eigenstates $\red{\ket{\psi}}$ are obtained by sums over symmetric/antisymmetric permutations \red{(Einstein summation convention is employed.)}
\begin{equation}
\red{\ket{\psi} = } t_{\alpha_1, \dots, \alpha_Q; \alpha_{Q+1} \dots \alpha_{2Q}; \alpha_{2Q+1} + \alpha_{3Q}} \ket{\alpha_1, \dots, \alpha_Q, \alpha_{Q+1} \dots \alpha_{2Q}, \alpha_{2Q+1} + \alpha_{3Q}}.
\end{equation}
We concentrate on the ground state, where the tensor has the following antisymmetry properties
\begin{align}
t_{\alpha_1, \alpha_2 \dots, \alpha_Q; \alpha_{Q+1} \dots \alpha_{2Q}; \alpha_{2Q+1} + \alpha_{3Q}} &= -t_{\alpha_2, \alpha_1 \dots, \alpha_Q; \alpha_{Q+1} \dots \alpha_{2Q}; \alpha_{2Q+1} + \alpha_{3Q}} &\text{ (Fermi-Dirac statistics within a given site)}\\
t_{\alpha_1, \alpha_2 \dots, \alpha_Q; \alpha_{Q+1} \dots \alpha_{2Q}; \alpha_{2Q+1} + \alpha_{3Q}} &= -t_{\alpha_{Q+1}, \dots, \alpha_Q; \alpha_{1} \dots \alpha_{2Q}; \alpha_{2Q+1} + \alpha_{3Q}} &\text{ ($H_H$ favors pairwise antisymmetry across sites)}
\end{align}
\end{widetext}

To get the total number of states, we start by overcounting allowed possibilities. There are $N$ options to place $\alpha_1$, $N-1$ to place $\alpha_2$ etc., leading to  
\begin{equation}
\frac{N !}{(N-3Q)!}
\end{equation}
states. However, we overcounted $3Q!$ different permutations, so the actual number of states is just
\begin{equation}
\left  (\begin{array}{c}
N \\ 
3Q
\end{array}  \right) = \left  (\begin{array}{c}
N \\ 
N-1
\end{array}  \right) = N.
\end{equation}

Thus, the following completely antisymmetriezed eigenstates are the ground state of the triangle at filling $Q$
\begin{equation}
 \ket{\alpha_N} = \frac{\epsilon_{\alpha_1 \dots \alpha_N}}{\sqrt{\mathcal N}} \ket{\alpha_1\dots \alpha_{Q};\alpha_{Q+1}\dots \alpha_{2Q}; \alpha_{2Q+1}\dots \alpha_{(N-1)} }. \label{eq:Groundstates}
 \end{equation} 

Here and in the following we label numerical normalization factors by $\mathcal N$. \red{This concludes the derivation of Eq.~\eqref{eq:TriangleGroundState}.} \EJK{There we use the notation
\begin{align}
&\ket{\text{singlet},m = 1} =\frac{\epsilon_{\alpha_1 \dots \alpha_N}}{\sqrt{\mathcal N}} \times \notag\\
& \quad \ket{\alpha_1\dots \alpha_{Q+1};\alpha_{Q+2}\dots \alpha_{2Q+1}; \alpha_{2Q+2}\dots \alpha_N},
\end{align}
and analogously for $m = 2, 3$.}


\subsection{Effective low-energy Hamiltonian}\label{sec:lowenergy}

As a next step, we project the Kondo-triangle Hamiltonian onto the groundstate manifold spanned by the $N$ states Eq.~\eqref{eq:Groundstates}. We begin by determining the spin-representation within the manifold of states Eq.~\eqref{eq:Groundstates}
\begin{widetext}
\begin{align}
\braket{\alpha_N \vert \hat{S}^a_m \vert \alpha_{N}'} = &\frac{\sigma^a_{\beta \beta'}\epsilon_{\alpha_1\dots \alpha_N} \epsilon_{\alpha_1' \dots \alpha_{N}'}}{\mathcal N}  \notag \\ &
\braket{\alpha_1\dots \alpha_{Q};\alpha_{Q+1}\dots \alpha_{2Q}; \alpha_{2Q+1}\dots \alpha_{N-1} \vert \underbrace{f^\dagger_{m,\beta} f_{m,\beta'}}_{\delta_{\beta  \beta'} - f_{m,\beta'} f^\dagger_{m,\beta}} \vert  \alpha_1'\dots \alpha_{Q'};\alpha_{Q+1}'\dots \alpha_{2Q}'; \alpha_{2Q+1}'\dots \alpha_{(N-1)}'} \notag \\
= &-\tilde{\mathcal N} \sigma^a_{\alpha_{N}' \alpha_N}.
\end{align}
\end{widetext}
This result immediately follows from the consideration that all spin quantum numbers except $\alpha_N (\alpha_N')$ have been used in the ket (bra). Thus the index of the creation operator $\beta = \alpha_N'$ ($\beta' = \alpha_N$) unless $\beta = \beta '$. We further used $\tr[\sigma^a ]= 0$.
Instead of explicitly calculating the positive proportionality constant we show that $\tilde{\mathcal N} = 1$ by 
\begin{align}
\sum_{N, N'} \vert \braket{\alpha_N \vert \hat{S}^a_m \vert \alpha_N'} \vert^2 &= \tr[ [\sigma^a]^2] \notag \\
&= \tilde{\mathcal N}^2 \tr[ \{[\sigma^a]^T\}^2].
\end{align}
\red{Here, the first equality follows from the completeness of} $\{ \ket{\alpha_N}\}$ and the second equality from the evaluation of the matrix element.
Therefore, the effective Hamiltonian has the form
\begin{equation}
H_{\rm eff} = H_c - \frac{J_K}{N}\sum_{m = 1}^3 [\hat{S}^a]^T c^\dagger_m \sigma^a c_m.
\end{equation}
As a final step, we reverse particle and hole operators $c_m \rightarrow h_m^\dagger, c_m^\dagger \rightarrow h_m$, then 
\begin{equation}
H_{\rm eff} = H_c + \frac{J_K}{N}\sum_{m = 1}^3 [\hat{S}^a]^T h^\dagger_m [\sigma^a]^T h_m.
\end{equation}

This is the origin of Eq.~\eqref{eq:Heff3CK} in the main text. \red{To see that Eq.~\eqref{eq:Heff3CK}, in which spin operators are transposed, is equivalent to the standard three channel Kondo model}
\begin{equation}
H_{\rm 3CK} = H_c + \frac{J_K}{N}\sum_{m = 1}^3 \hat{S}^a h^\dagger_m \sigma^a h_m,
\end{equation}
\red{it is sufficient to realize that the SU(N)-invariant interaction can be reexpressed using the Fierz identity}
\begin{equation}
[\sigma^a]_{\alpha \beta} [\sigma^a]_{\gamma \delta} = \delta_{\alpha \delta} \delta_{\beta \gamma} - \frac{1}{N} \delta_{\alpha \beta} \delta_{\gamma \delta},
\end{equation}
\red{which is invariant under simultaneous transposition operation $(\alpha, \gamma) \leftrightarrow (\beta, \delta)$.}

\subsection{Robustness against inhomogeneity}

At strong coupling the triangle is robust against moderate inhomogeneities in $J_H$, as can be seen by the following evaluation of matrix elements of $\delta H =  \delta J_H \hat{S}_1^a \hat{S}_2^a$
\begin{align}
\braket{\alpha_N \vert \delta H \vert \alpha_N'} &= \delta J_H \sum_{\tilde \alpha_N} \braket{\alpha_N \vert \hat{S}_1^a \vert \tilde \alpha_N}\braket{\tilde \alpha_N \vert \hat{S}_2^a \vert \alpha_N'} \notag \\
&= \tilde{\mathcal N}^2 \sum_a\sum_{\tilde \alpha_N} [\sigma^a]_{\alpha_N' \tilde \alpha_N} [\sigma^a]_{\tilde \alpha_N \alpha_N} \propto \delta_{\alpha_N' \alpha_N}.
\end{align}
Thus, inhomogeneities projected to the ground state manifold are proportional to the unit matrix and do not lift the degeneracy of states $\ket{\alpha_N}$.

\section{Impurity partition sum, static evaluation}
\label{app:ImpurityPartition}

In this section we present technical details on the evaluation of the partition sum. Throughout the paper, we consider the partition sum (and thus free energy and effective action) of the impurity alone. This is defined as $\mathcal Z_{\rm impurity} = \mathcal Z_{\rm total}/\mathcal Z_{\rm no\; impurity}$, where $Z_{\rm total}$ is given by 
\begin{widetext}
\begin{equation}
\mathcal Z_{\rm total} = \prod_{m = 1}^3 \int_0^\infty\mathcal D V_m V_m \int_{-i\infty}^{i\infty} \mathcal D \lambda_m \int_{\mathbb C^2} \mathcal D[t_m, t_m^*] \int \mathcal D[c_m,f_m]e^{- S[V_m,\lambda_m, t_m, c_m, f_m]}, \label{eq:ReadNewns}
\end{equation}
\end{widetext}

Note that we employ Read-Newns gauge ($V_m>0$) throughout this section.
The partition sum $\mathcal Z_{\rm no\; impurity}$ is the same partition sum of the three wires but without any Kondo impurities.

\subsection{Diagonalization of spinon Hamiltonian}
\label{sec:DiagonalizationHam}

The spinon Hamiltonian, see also Eq.~\eqref{eq:AbrikosovAction} of the main text, has the form
\begin{equation}
H_t = - t f^\dagger \left (\begin{array}{ccc}
0 & e^{i A_1} & e^{-i A_3} \\ 
e^{-i A_1} & 0 & e^{i A_2} \\ 
e^{i A_3} & e^{-i A_2} & 0
\end{array} \right ) f,\label{eq:Ht}
\end{equation}
where we use the three component notation $f = (f_1, f_2, f_3)$, and similarly for $c(\v x) = (c_1 (\v x), c_2(\v x), c_3(\v x))$ on each site of the wires.
We rotate $f  = U \tilde f$ and $c = U \tilde c$ electrons by $U = \text{diag}(e^{i (A_1 - \Phi/3)},1,e^{-i (A_2 - \Phi/3)})$, leading to 
\begin{equation}
H_t = - t \tilde f^\dagger \left (\begin{array}{ccc}
0 & e^{i \Phi/3} & e^{-i \Phi/3} \\ 
e^{-i \Phi/3} & 0 & e^{i\Phi/3} \\ 
e^{i \Phi/3} & e^{-i \Phi/3} & 0
\end{array} \right ) \tilde f.
\end{equation}
This rotation appears at the expense of a vector potential
\begin{equation}
(c^\dagger, f^\dagger) \partial_\tau (c,f)^T = (\tilde c^\dagger, \tilde f^\dagger) [\partial_\tau + i \mathcal A] (\tilde c,\tilde f)^T,
\end{equation}
where
\begin{equation}
\mathcal A = -i U^\dagger \partial_\tau U = \text{diag}(\dot A_1 - \dot \Phi/3,0, - [\dot A_2- \dot \Phi/3]).
\end{equation}
It is furthermore useful to expand $\tilde f, \tilde c$ in eigenstates with instantaneous energy $\epsilon_k = -2t \cos(k + \Phi(\tau)/3)$
\begin{equation}
\ket{\psi_k} = \frac{1}{\sqrt{3}}\left (\begin{array}{c}
e^{- i k} \\ 
1 \\ 
e^{ i k}
\end{array} \right ), \quad k = 0, \pm \frac{2\pi}{3}.
\end{equation}
In this basis the Berry connection is $
\mathcal A_{k'k} = ({[\dot A_1- \dot \Phi/3] e^{i(k'- k)} - [\dot A_2}- \dot \Phi/3]e^{-i(k'- k)})/{3}$. In summary, the total Lagrangian under consideration is (we employ the notation $D_\tau = \partial_\tau + i \mathcal A$ and $\lambda_k =  \lambda + \epsilon_k(\Phi)$)
\begin{widetext}
\begin{eqnarray}
\mathcal L &=& \sum_{k,k'} \left(\begin{array}{cccc}
\cdots & c^\dagger_{\alpha,k}(\v p) & \cdots & f^\dagger_{\alpha,k} 
\end{array} \right) \left (\begin{array}{ccc|c}
 &  &  &  \\ 
 & [D_\tau]_{k,k'}+\epsilon(\v p) \delta_{\v p, \v p'} \delta_{kk'}  &  & V \delta_{kk'}\\ 
 &  &  &  \\ 
\hline
 & V\delta_{kk'} &  & [D_\tau]_{k,k'} +\lambda_k \delta_{kk'}
\end{array}  \right) \left (\begin{array}{c}
\vdots \\ 
c_{\alpha,k'}(\v p')\\
\vdots \\
f_{\alpha,k'}
\end{array} \right)\notag \\
&& + 3 \left (N \frac{t^2}{J_K} + N \frac{V^2}{J_K} - \lambda q N \right ). \label{eq:LTOTappendix}
\end{eqnarray}
\end{widetext}

\subsection{Static fields and mean field solution}
\label{app:MeanField}
We begin by studying the mean field solution.
At this level, we consider all bosonic fields $V>0, t>0,\Phi = \sum_m A_m$ as constant variational parameters, and $\mathcal A = 0$. The fermionic integral yields Eq.~\eqref{eq:FermionDeterminant}.
The mean field equations involve the following two integrals
\begin{subequations}
\label{eq:EqualTimeGF}
\begin{eqnarray}
n_F \equiv I_1(\lambda/\Delta) &=& T \sum_n \frac{e^{i \epsilon_n \eta}}{i\epsilon_n - \lambda + i \Delta s(\epsilon_n/D)} \notag \\
&\simeq& \text{arccot}(\lambda/\Delta)/\pi  =: \delta(\lambda/\Delta)/\pi,\\
I_2(\lambda + i \Delta) &=& T \sum_n \frac{is(\epsilon_n/D) e^{i \epsilon_n \eta}}{i\epsilon_n - \lambda + i \Delta s(\epsilon_n/D)} \notag \\
&\simeq & -\frac{\ln(\vert \lambda + i \Delta \vert \eta) + \gamma_{\rm EM}}{\pi}.
\end{eqnarray}
\end{subequations}
Here, $\gamma_{\rm EM}$ is the Euler Mascheroni constant (with our regularization scheme $T_K = e^{-1/[\rho J_K] - \gamma_{\rm EM}} /\eta$) and $\simeq$ implies a zero temperature calculation. Note that $\delta$ becomes a step function (from $\pi$ down to $0$) as $\Delta \rightarrow 0$.  

Having established these prerequisites, we are now in the position of imposing the mean field equations
\begin{subequations}
\begin{eqnarray}
\frac{1}{N} \frac{\partial F}{\partial \lambda} = \sum_k \left ( I_1 \left (\frac{\lambda_k}{\Delta} \right) - q \right) \stackrel{!}{=} 0, \\
\frac{1}{N} \frac{\partial F}{\partial \Delta} = \sum_k \left ( -I_2 \left (\frac{\lambda_k}{\Delta} \right) +\frac{1}{\pi \rho J_K} \right) \stackrel{!}{=} 0, \\
\frac{1}{N} \frac{\partial F}{\partial t} = \sum_k \left ( \frac{\partial \epsilon_k}{\partial t}I_1\left (\frac{\lambda_k}{\Delta} \right) +\frac{2 t}{J_H} \right) \stackrel{!}{=} 0, \\
\frac{1}{N} \frac{\partial F}{\partial \Phi} =\sum_k \frac{\partial \epsilon_k}{\partial \Phi}I_1\left (\frac{\lambda_k}{\Delta} \right) \stackrel{!}{=} 0.
\end{eqnarray}
\end{subequations}
We readily see that ground state solutions are given by $\Phi \in 2\pi \mathbb Z$. Then the first three equations yield \red{cf. Eqs.~\eqref{eq:MFCondAll},}
\begin{widetext}
\begin{subequations}
\begin{align}
3 \pi q &= \sum_k \delta_k  = \delta_{\red{0}} + 2 \delta_{\red{2\pi/3}} \\
T_K^3 &= \prod_k \sqrt{\lambda_k^2 + \Delta^2} = \frac{\Delta^3}{\sin(\delta_3) \sin(\delta_{\red{2\pi/3}})^2} \\
2( \delta_{\red{0}} - \delta_{\red{2\pi/3}})&=3 \frac{2\pi t}{J_H} = \frac{2\pi }{J_H} (\lambda_{\red{2\pi/3}} - \lambda_{\red{0}}) =  \frac{2\pi \Delta}{J_H} (\cot(\delta_{\red{2\pi/3}}) -  \cot(\delta_{\red{0}})) = 2 \sin(\delta_{\red{0}} - \delta_{\red{2\pi/3}}) \frac{\pi T_K}{J_H} \sqrt[3]{\frac{1}{\sin(\delta_{\red{2\pi/3}}) \sin^2 (\delta_{\red{0}})}} 
\end{align}
\label{eq:MeanFieldEq}
\end{subequations}
\end{widetext}
 We readily recognize the Kondo solution $t = 0$, $\delta_k = \pi q, \Delta = T_K \sin(\pi q)$, which is present for any $T_K/J_H$. \red{The last equation is the origin of Eq.~\eqref{eq:MFCond}}.
 


\subsection{Finite temperatures}
\label{app:FiniteTempMeanField}

Of the presented finite temperature phases in Fig.~\ref{fig:Schematics} {\bf b} of the main text, the presence of the LFL and local moment phase is obvious. The existence of an LFL$^*$ and of the 3CK phase is discussed now by showing that there is a mean field transition $T_{\rm SL} = J_H q (1-q)$ below which $t$ develops a vacuum expectation value and a lower transition $T_{K}^{\rm eff}$ at which $V$ spontaneously develops. 
For the perturbative solution in $\Delta$ at finite $T$ we use 
\begin{equation}
n_f (\lambda) = I_1 = n_{\rm FD} (\lambda) = \frac{1 - \tanh(\lambda/2T)}{2}
\end{equation}
and, perturbatively in $\Delta$,
\begin{eqnarray}
I_2(\lambda) &=& T \sum_{\epsilon_n>0} \left (\frac{i}{i \epsilon_n - \lambda} + \frac{i}{i \epsilon_n + \lambda}\right) \notag \\
& \simeq& \frac{\ln(D/T)}{\pi} -\frac{\psi ^{(0)}\left(\frac{i \lambda/T +\pi }{2 \pi }\right)+\psi ^{(0)}\left(\frac{\pi -i \lambda/T }{2 \pi }\right)}{2 \pi }.
\end{eqnarray}

The mean field equations (perturbative in $\Delta$) are then
\begin{widetext}
\begin{eqnarray}
  n_{\rm FD}(\lambda -2t) &=& q +\frac{2t}{J_H} = q + \frac{\lambda}{J_H} - \frac{\lambda - 2 t}{J_H},\\
  n_{\rm FD}(\lambda +t) &=& q - \frac{t}{J_H}, \\
3 \ln\left ( \frac{T_K}{T} \right ) &=& \sum_k \left ( \frac{\psi ^{(0)}\left(\frac{i \lambda_k/T +\pi }{2 \pi }\right)+\psi ^{(0)}\left(\frac{\pi -i \lambda_k/T }{2 \pi }\right)}{2} -   \frac{\psi ^{(0)}\left(\frac{i \lambda_{t=0}/T +\pi }{2 \pi }\right)+\psi ^{(0)}\left(\frac{\pi -i \lambda_{t=0}/T }{2 \pi .}\right)}{2} \right), \label{eq:KondoFiniteT}
\end{eqnarray}
\end{widetext}
with $\lambda_k = (\lambda + t, \lambda + t, \lambda - 2t)$ and $\lambda_{t=0} = 2 T \text{artanh}(1 -2q)$ the solution without $t$.
The mean field spin-liquid transition temperature is obtained by expanding the first two equations in $t$
\begin{eqnarray}
q &=& n_{\rm FD}(\lambda) \Leftrightarrow \lambda = 2T \text{artanh} \left (1 - 2 q \right),\\
\frac{1}{J_H} &=& -\frac{\partial n_{\rm FD}}{\partial \lambda} = \frac{1}{4T \cosh^2(\lambda/2T)} \notag \\
 &=& \frac{1}{4T \cosh^2(\text{artanh}(1- 2q))} .
\end{eqnarray}
Thus for $0<q<1/3$
\begin{equation}
{\frac{T_{\rm SL}}{J_H} = \frac{1}{4\pi \cosh^2(\text{artanh}(1- 2q))} = q (1 - q) .}
\end{equation}
For the solution of $T_K^{\rm eff} < T_{\rm SL}$ it is more convenient to use 
\begin{equation}
n_{\red{k}} = n_{\rm FD} (\lambda_{\red{k}})
\end{equation}
and insert this into 
\begin{eqnarray}
3 q &=& 2n_{\red{2\pi/3}} + n_{\red{0}}, \\
\Delta n  \equiv n_{\red{0}} - n_{\red{2\pi/3}} &=& \frac{3 t}{J_H} = \frac{T}{J_H} (\bar \lambda_{\red{2\pi/3}} - \bar \lambda_{\red{0}}).
\end{eqnarray}
We use 
\begin{eqnarray}
\bar \lambda_{t = 0} = \lambda_{t = 0}/T &=& 2  \text{artanh}(1 - 2 q), \\
\bar \lambda_{\red{2\pi/3}} = \lambda_{\red{2\pi/3}}/T&=& 2 \text{artanh}(1 - 2 n_{\red{2\pi/3}}) \notag \\
&=& 2 \text{artanh}(1 - 2 q + \frac{2 \Delta n}{3}), \\
\bar \lambda_{\red{0}}= \lambda_{\red{0}}/T &=& 2 \text{artanh}(1 - 2 n_{\red{0}}) \notag \\
&=& 2 \text{artanh}(1 - 2 q - \frac{4 \Delta n}{3}).
\end{eqnarray}
to replace temperature in Eq.~\eqref{eq:KondoFiniteT}
\begin{widetext}
\begin{align}
T &= T_K \prod_k \left [\text{exp} \left ( \frac{\psi ^{(0)}\left(\frac{i \lambda_k/T +\pi }{2 \pi }\right)+\psi ^{(0)}\left(\frac{\pi -i \lambda_k/T }{2 \pi }\right)}{2} -   \frac{\psi ^{(0)}\left(\frac{i \lambda_{t=0}/T +\pi }{2 \pi }\right)+\psi ^{(0)}\left(\frac{\pi -i \lambda_{t=0}/T }{2 \pi }\right)}{2} \right) \right]^{-1/3} \notag \\ 
&\equiv T_K \underbrace{g(\lambda_{t=0},\lambda_{\red{2\pi/3}}(\Delta n),\lambda_{\red{0}}(\Delta n))}_{:=f(\Delta n)}.
\end{align}
\end{widetext}
We thus reduced the finite temperature Kondo transition in the presence of finite $t$, i.e.~finite $\Delta n$ to a single equation for $\Delta n$
\begin{equation}
{\Delta n = \frac{T_K}{J_H} f(\Delta n) [\bar \lambda_{\red{2\pi/3}} - \bar \lambda_{\red{0}}].}
\end{equation}
Numerical solution of this equation demonstrates the existence of $0<T_{K}^{\rm eff}<T_{\rm SL}$ for sufficiently small $T_K/J_H$.

\subsection{Landau Free energy (perturbative in t)}
\label{app:Landau}

We consider the case of small $t$ and employ $\xi = \lambda + i \Delta = T_K e^{i \pi q}$~\cite{Coleman2015}
\begin{eqnarray}
V[\Phi] &=& \frac{N}{\pi} \sum_k \text{Im}\left [ (\lambda_k + i \Delta)\ln\left (\frac{(\lambda_k + i \Delta)}{e T_K e^{i \pi q}}\right )\right] \\
&=& \frac{N}{\pi} \sum_k \text{Im} \left [ \sum_k  \frac{\epsilon_k^2}{2\xi} - \frac{\epsilon_k^3}{6\xi^2} + \frac{\epsilon_k^4}{12\xi^3}\right] \notag \\
&=& \frac{T_K N}{\pi} \Big [- 3 \bar t^2 \sin(\pi q) - \cos(\Phi) \bar t^3 \sin( 2\pi q)\notag \\
&& -3 \bar t^4 \sin(3\pi q)/2\Big ].
\end{eqnarray}

Up to the effect of biquadratic and ring exchange terms (see following section), as well as the Hubbard-Stratonovich term $3t^2/J_H$, this expression yields Eq.~\eqref{eq:Landau} of the main text.

\subsection{Ring exchange and biquadratic terms}

In the large $N$ limit, the transition between LFL  and 3CK appears to be first order. Here, we consider additional terms which ultimately overcome the first order behavior. We need 
\begin{equation}
\langle f_m f^\dagger_{m+1} \rangle \simeq - \int (d \epsilon) \frac{t_m}{[i \epsilon + \lambda + i \Delta \text{sign}(\epsilon)]^2}  = \frac{t_m}{\pi T_K} \sin(\pi q) 
\end{equation}

We first study ring-exchange terms of the form
\begin{equation}
H_3 = -\pi^2\frac{J_s}{N} d_{abc} \hat{S}_1^a\hat{S}_2^b\hat{S}_3^c - \pi^2\frac{J_\chi}{N} f_{abc} \hat{S}_1^a\hat{S}_2^b\hat{S}_3^c.
\end{equation}

These terms can be evaluated on mean field level as ($T_{abc} = J_s d_{abc} + J_\chi f_{abc}$ and 
we use $t_{m} = t  e^{i \Phi/3}$.)
\begin{align}
H_3 &\simeq  -\pi^2\frac{T_{abc}}{N} \langle f^\dagger_1 \sigma^a f_1 f^\dagger_2 \sigma^b f_2f^\dagger_3 \sigma^c f_3 \rangle  \notag \\
&=\frac{T_{abc} (\sin \pi q)^3}{\pi T_K^3 N}  [t_{1} t_{2} t_{3} \tr(\sigma^a \sigma^b \sigma^c) +\bar t_{3}  \bar t_{2} \bar t_{1} \tr(\sigma^a \sigma^c \sigma^b)]  \notag\\
&= J_s (\sin \pi q)^3\frac{d_{abc}}{\pi T_K^3 N}\underbrace{\tr(\sigma^a \lbrace \sigma^b, \sigma^c\rbrace)}_{d_{abc}} t^3 \cos(\Phi) \notag \\
&+  J_\chi (\sin \pi q)^3 \frac{if_{abc}}{\pi T_K^3 N}\underbrace{\tr(\sigma^a [\sigma^b, \sigma^c])}_{if_{abc}}t^3 \sin(\Phi) \notag \\
&= N\frac{J_s (\sin \pi q)^3}{\pi T_K^3} t^3 \cos(\Phi) +  N  \frac{J_\chi (\sin \pi q)^3}{\pi T_K^3}t^3 \sin(\Phi) 
\end{align}
We used
\begin{equation}
d_{abc} d_{abc} = N^2 - 4,\; f_{abc} f_{abc} = N^2.
\end{equation}
This term enters $\beta$ in Eq.~\eqref{eq:Landau} of the main text.

We furthermore introduce biquadratic interactions
\begin{equation}
H_4 = -\frac{\pi^3J_4}{2N^3} \sum_m [\hat{S}_m^a \hat{S}_{m+1}^a]^2.
\end{equation}
Their mean field decoupling leads to 
\begin{eqnarray}
H_4 &=& -\frac{\pi^3J_4}{2N^3} \llangle \sum_m f_m^\dagger \sigma^a f_m f_{m+1}^\dagger \sigma^a f_{m+1} f_m^\dagger \sigma^b f_m f_{m+1}^\dagger \sigma^b f_{m+1} \rrangle \notag \\
&=& 3t^4\frac{J_4 \sin(\pi q)^4}{2N^3 \pi T_K^4} \left (\underbrace{\tr[\sigma^a \lbrace \sigma^a, \sigma^b \rbrace \sigma^b]}_{d_{abc} d_{abc}/2} + \underbrace{\tr[\sigma^a \sigma^a]^2}_{=(N^2-1)^2} \right )\notag \\
&\simeq& N \frac{3 J_4 \sin(\pi q)^4}{2\pi T_K^4} t^4.
\end{eqnarray}
This term enters $\gamma$ in Eq.~\eqref{eq:Landau} of the main text.

For the plot of Fig.~\ref{fig:Goldstone} {\bf c} we used $J_{\rm Ring}^{\rm eff} = 0.3$ and $J_4^{\rm eff} = 0.1$ in the effective replacement $J_H \rightarrow J_H(1 + d J_{\rm ring}^{\rm eff} - d^2 J_4^{\rm eff})$ in the numerator of Eq.~\eqref{eq:MFCond}, left. 

The replacement \red{is related to the microscopic Hamiltonian as} follows. From the mean field evaluation 
\begin{equation}
\langle f_m f_{m+1}^\dagger \rangle = \sum_k G_k(0)/3/3 = - d/3\pi.
\end{equation}
Therefore, on Hartree-Fock level
\begin{eqnarray}
H_3 &\rightarrow& -N J_s (d/3)^3/\pi = N \pi^2 J_s (t/J_H)^3, \\
 \quad H_4 &\rightarrow& N J_4 3 (d/3)^4/(2\pi) = N J_4 3 \pi^3 (t/J_H)^4/2.
\end{eqnarray}
For small $t/J_H$ \red{this} can be reinterpreted as a renormalization 
\begin{eqnarray}
\frac{t^2}{J_H} &\rightarrow& \frac{t^2}{J_H (1 - \pi^2 J_s t/J_H^2 - 3 \pi^2 (t^2/J_H^3) J_4/2)} \notag \\
&=&\frac{t^2}{J_H (1 + \pi J_s/J_H\, d/3 - J_4/J_H\, d^2/3)}.
\end{eqnarray}
Hence we identify
\begin{equation}
J_{\rm Ring}^{\rm eff} = \frac{\pi J_s}{3 J_H}, \; J_{\rm 4}^{\rm eff} = \frac{J_4}{3 J_H} .
\end{equation}

%

\section{Dynamics of Goldstone modes and total flux}
\label{app:Fluctuations}
In this section we derive the kinetic terms for Goldstone bosons and $\Phi(\tau)$, \red{Eq.~\eqref{eq:SA} of the main text}. 

\subsection{Goldstone bosons}
\label{app:DerivGoldstone}

Before turning to the effective action of Goldstone bosons we comment on the structure
\begin{equation}
\frac{G}{H} = \frac{U(1) \times U(1) \times U(1)}{U(1) \times \mathbb Z_3}
\end{equation}
of the Goldstone manifold. Smooth transformations of the large group $G$ are represented by three phases $\chi_m$ 
\begin{equation}
f_m \rightarrow U_{mm'} f_{m'}; \quad c_m \rightarrow U_{mm'} c_{m'} \quad (U_{mm'} = \delta_{mm'} e^{i \chi_m})
\end{equation}
and per definition $\chi_m (\tau = 1/T) = \chi_m(\tau = 0) + 2\pi j$. Following Sec.~II of this supplement, a convenient form of $U = \text{diag}(e^{i A_1 - \Phi/3},1,e^{-i A_2 + \Phi/3})$, as it cancels the fluctuating gauge fields on the links. To make the quotient group $G/H$ apparent we factorize 
\begin{equation}
U = e^{i \phi} V, \text{ with } \det V = 1.
\end{equation}
The naive derivation of the Goldstone action implies the absorption the $V$ (i.e.~the $SU(3)$ part of $G$) into $f, c$ at the expense of a Berry curvature term $\mathcal A = - i V^\dagger \partial_\tau V$. The integration of fermions then leads to an effective action in terms of $\mathcal A$, an thus implicitly in terms of $A_{1,2,3}$.

However, a certain care is needed for this procedure. The quotient group introduces an emergent $\mathbb Z_3$ redundancy which is manifested in non-contractable loops $(j = 0,1,2)$
\begin{eqnarray}
e^{i\phi(1/T)} &=& \omega^j e^{i\phi(0)}, \\
V(1/T) &=& \bar \omega^j V(0).
\end{eqnarray}
In particular, the absorption of $f(\tau) = V(\tau)\tilde f(\tau)$ changes the boundary conditions ($\tilde f(1/T) = - \omega^j f(0)$), i.e.~$\tilde f(\tau)$ is generically not a fermionic field. We conjecture that the topological nature of $\pi_1(G/H) = \mathbb Z_3$ is at the root of the ground state degeneracy of the 3CK phase.

To remedy this problem we choose a parametrization of $U(\tau)$ such that the topological winding is manifest, i.e.
\begin{eqnarray}
U(\tau) &=& e^{i \bar \phi(\tau) + 2 \pi i \tau T/3} \notag \\
&&\times \left (\begin{array}{ccc}
e^{-2 \pi i \tau T/3} & 0 & 0 \\ 
0 & e^{-2 \pi i \tau T/3} & 0 \\ 
0 & 0 & e^{4 \pi i \tau T/3}
\end{array} \right ) \bar V(\tau),
\end{eqnarray}
where both $e^{i \bar \phi(\tau)}$ and $\bar V(\tau)$ are periodic in imaginary time. In this parametrization it is apparent that the three different $\mathbb Z_3$ sector correspond to the $2\pi$ winding of one of the $\chi_m$. In order to derive the effective action of $V(\tau)$ fluctuations even for non-zero $j$, we thus absorb $e^{- i \bar \phi(\tau)} U(\tau)$ into fermionic fields (without changing their statistics) and integrate fermions subsequently.
%
%
%
(with $G_{cf}, \red{G_{cf,k}}$ the full Green's function of c and f space and $G_c, \red{G_{c,k}}$ the Green's function of conduction electrons \red{and $G, G_k$ the Green's function of $f$ electrons})
\begin{eqnarray}
S[A_{2,3}]/N &=& -  \Tr \ln [- G_{cf}^{-1} + i \mathcal A] +  \tr \ln [- G_{c}^{-1} + i \mathcal A] \\
&\simeq & i  \Tr [G_{cf} \mathcal A] - i  \tr [G_{c} \mathcal A] \notag \\
&& - \frac{1}{2} \Tr[( G_{cf} \mathcal A)^2] + \frac{1}{2} \tr[( G_{c} \mathcal A)^2].
\end{eqnarray}
The symbol ``$\tr$'' denotes a trace in the space of the three sites and in time, ``$\Tr$'' additionally includes the $2\times 2$ space of c and f electrons. Specifically, we employ a gauge in which
\begin{equation}
\mathcal A_{k'k} = \frac{2\pi j T}{3} + \underbrace{\frac{(\dot A_2 - \dot A_1) \delta_{kk'} + \left (\dot A_1 e^{i (k' - k)}- \dot A_2 e^{-i (k' - k)}\right )}{3}}_{=:\bar{\mathcal A}_{k'k}}.
\end{equation}

The leading term is fixed by the constraint $\sum_k \delta_k = 3\pi q$ (this result is true beyond mean field)
\begin{eqnarray}
S^{(1)} &=& i \int d \tau \sum_k G_{k} (\tau, \tau^+) \mathcal A_{kk}(\tau) \notag \\
&=& i Q \int d\tau \sum_k\mathcal A_{kk}(\tau)\notag \\
&=& i Q 2\pi m.
\end{eqnarray}
Note that, since $Q \in \mathbb Z$, this expression is invariant yields a trivial phase $2\pi$ and can be omitted.

Next we switch to the term of second order in gradients, which can be expressed as 
\begin{equation}
S^{(2)} = -\frac{N}{2} \int d\tau \sum_{kk'}I_{kk'} \vert \mathcal A_{kk'} \vert^2 . \label{eq:S2}
\end{equation}
The polarization operator under consideration is
\begin{align}
I_{kk'} &= \int \frac{d \epsilon}{2\pi} \tr^{\rm cf}[G_{cf , k}(\epsilon) G_{cf , k'}(\epsilon)] -G_{c , k}(\epsilon) G_{c , k'}(\epsilon) \notag\\
&= (1 + 2 (\pi \rho V)^2) \int \frac{d \epsilon}{2\pi} G_{ k}(\epsilon) G_{ k'}(\epsilon) \notag \\
&= - \frac{1 + 2 \rho \Delta}{J_H} \left (\begin{array}{ccc}
\frac{J_H \sin(\delta_+)^2}{\Delta \pi} & \frac{J_H \sin(\delta_+)^2}{\Delta \pi} & 1 \\ 
\frac{J_H \sin(\delta_+)^2}{\Delta \pi} & \frac{J_H \sin(\delta_+)^2}{\Delta \pi} & 1 \\ 
1 & 1 & \frac{J_H \sin(\delta_0)^2}{\Delta \pi}
\end{array} \right )_{kk'}.
\label{eq:Ikkp}
\end{align}
We here used \red{the short hand} notation $\delta_+ = \delta_{2\pi/3}$ and that
\begin{widetext}
\begin{equation}
 \int \frac{d \epsilon}{2\pi} G_{ k}(\epsilon) G_{k'}(\epsilon) =- \frac{1}{\pi} \begin{cases} \frac{\Delta}{\Delta^2 + \lambda_k^2}, & \lambda_k = \lambda_{k'} \\
 \frac{\pi + \arctan(\Delta/\lambda_k) - \arctan(\Delta/\lambda_{k'})}{\lambda_{k'} -\lambda_k}, & \lambda_k <0, \lambda_{k'} >0,\end{cases}
\end{equation}
\end{widetext}
as well as the mean field equations, Eq.~\eqref{eq:MeanFieldEq}.
It is important to realize that remnant $U(1)$ terms and $SU(3)$ terms in Eq.~\eqref{eq:S2} decouple
\begin{equation}
S^{(2)} =  -\frac{N}{2} \int d\tau \sum_{kk'} [I_{kk'} \vert \bar{\mathcal A}_{kk'} \vert^2 + I_{kk} \left( \frac{2\pi m T}{3 }\right)^2 \delta_{kk'} ]. \label{eq:S2GaugeTrafo}
\end{equation}

The second term yields a vanishing contribution to the weight in the limit $T \rightarrow 0$ and is disregarded. We further use that
\begin{equation} \label{eq:Asquared}
\vert \bar{\mathcal A}_{kk'} \vert^2  = \frac{\sum_m \dot A_m^2}{18} .
\end{equation} 
\red{Here, we have used a gauge transformation to return to generic gauge. The combination of Eq.~\eqref{eq:Ikkp}, \eqref{eq:S2GaugeTrafo}, \eqref{eq:Asquared}, results in the final result, Eq.~\eqref{eq:GaugeAction} in the main text.
To obtain Eq.~\eqref{eq:GoldstoneAction}}, we employ the parametrization in terms of unit vectors $\hat e_m$ and $\sum_m \hat e_m \hat e_m^T = 3 \mathbf 1$.

\subsection{Total flux}
\label{app:SDissDeriv}
To obtain the dynamics of the total flux $\Phi(\tau)$ we use \red{the notation} 
$
\delta \lambda_k = - 2 t \left [ \cos(k + \Phi/3) - \cos(k)  \right] 
$
and \red{expand the fermionic determinant to} 
second order 
in $\delta \lambda_k$. 
\begin{widetext}
\begin{eqnarray}
S_{\rm eff}^{(2)} &\simeq& \frac{N}{2} \sum_{\omega_m,k} \delta\lambda_k(\omega_m)\delta\lambda_k(-\omega_m) \sum_{\epsilon_n}G_k(\epsilon_n) G_k(\epsilon_n + \omega_m)\notag \\
& \stackrel{T \rightarrow 0}{\simeq}& \frac{N {\Delta}}{2 \pi T} \sum_{\omega_m,k} \delta\lambda_k(\omega_m)\delta\lambda_k(-\omega_m)  \frac{\ln \left (\frac{\lambda_k^2 + \Delta^2}{\lambda_k^2 + (\Delta + \vert \omega \vert)^2}\right )}{\vert \omega \vert (2 \Delta + \vert \omega\vert)} \notag\\
&=&\frac{N\Delta}{2 \pi T} \sum_{\omega_m,k} \delta\lambda_k(\omega_m)\delta\lambda_k(-\omega_m) \begin{cases}\left [ - \frac{1}{\Delta^2 + \lambda_k^2} + \vert \omega \vert \frac{\Delta}{(\Delta^2 + \lambda^2_k)^2}\right ], & \vert \omega \vert \ll \lambda^2_k + \Delta^2, \\
 - \frac{\ln\left (\frac{\omega^2}{\lambda_k^2 + \Delta^2}\right)}{\omega^2}, & \vert \omega \vert \gg \lambda^2_k + \Delta^2. \end{cases}\label{eq:SPhi}
\end{eqnarray}
\end{widetext}
The equation substantially simplifies for small $t \ll \lambda$, and leads to the kinetic energy of $\Phi$ fluctuations
\begin{eqnarray}
S_{\rm kin}[\Phi] &=& N\frac{3 t^2 T_K \sin(\pi q)}{4 \pi T} \sum_{\omega_m} \Phi(\omega_m)\Phi(-\omega_m) \notag\\
&&\times \begin{cases} \vert \omega \vert \frac{\sin(\pi q)}{T_K^3}, & \vert \omega \vert \ll T_K^2, \\
\left [ \frac{1}{T_K^2} - \frac{\ln\left (\frac{\omega^2}{T_K^2}\right)}{\omega^2} \right ], & \vert \omega \vert \gg T_K^2. \end{cases}
\end{eqnarray}

The $\vert \omega \vert$ term in the first line is the origin of the damped kinetic term presented 
in the main text and leads to logarithmic correlators.

\subsection{Estimate of tunneling time and tunneling action.}
\label{app:Estimate}
As demonstrated in the main text, details of the tunneling rate $\Gamma$ are irrelevant for the transition. We therefore constrain ourselves to merely estimate $\Gamma$, based on a tunneling event
\begin{align}
\Phi(\tau) &= \frac{\pi}{2} + 2\pi \tau/\tau_0 \theta(\tau_0^2-4 \tau^2) \notag \\
\Rightarrow \vert\Phi(\omega)\vert^2 &= \frac{\left(\omega \tau_0  \cos \left(\frac{\omega \tau_0}{2}\right)-2 \sin \left(\frac{\omega\tau_0 }{2}\right)\right)^2}{[\omega\tau_0] ^4}.
\end{align}

In terms of dimensionless parameters $\bar t= t/T_K$ and $\bar \tau_0 = \tau_0/T_K, \bar \omega = \omega \tau_0, \Phi(\omega) = \bar \Phi(\omega \tau_0)$ we obtain
\begin{widetext}
\begin{eqnarray}
S_{\rm tun}(\bar \tau_0)/N &\sim& \beta \bar \tau_0 +\bar t^2 \sin(\pi q) \bar \tau_0 \int_0^\infty \vert \bar \Phi(\bar \omega) \vert^2 \left (1 -\frac{ \ln([\sin(\pi q)+\bar\omega]^2 \bar \tau_0^2)}{\bar \omega^2 \bar \tau_0^2} \right) \notag \\
&\approx & \beta \bar \tau_0 \left \lbrace 1+\frac{ \bar t^2 \sin(\pi q)  }{\beta} \left  [\frac{\pi}{6} - \frac{1}{\bar \tau_0^2} \left (\frac{\pi \ln[\sin(\pi q)^2 \bar \tau_0^2]}{60} + 0.7\right )\right ] \right \rbrace.
\end{eqnarray}
\end{widetext}
To obtain the optimal tunneling time we use that $\bar t \sim \beta/\gamma$ at the mean field first order transition. We thus obtain for any $q$ such that $\sin(\pi q) \sim 1$
\begin{eqnarray}
\tau_0^{\rm optimal} &\sim& \frac{1}{T_K}\begin{cases}
1, & \gamma \ll 1,\\ \sqrt{\frac{\ln(\gamma) }{\gamma}}, & \gamma \gg 1, \end{cases} \\
S_{\rm tun}/N &\sim & \beta \begin{cases}
1/\gamma + 1.4, & \gamma \ll 1.\\ \sqrt{\frac{\ln(\gamma) }{\gamma}}, & \gamma \gg 1. \end{cases}
\end{eqnarray}
as quoted in the main text.

\subsection{Implications for interwire correlations}
Using $\psi = (c,f)^T$ the generating functional at mean field level, but including fluctuations of the Goldstone modes is $Z = \prod_k Z_k$
\begin{align}
Z_k[\eta] &= \int \mathcal D \psi_k e^{- \int d \tau \bar \psi_k [- \hat G^{-1} + i \mathcal A] \psi_k + \bar\psi_k U\eta_k + \bar \eta_k U^\dagger \psi_k } \notag \\
&= e^{\int d \tau \bar \eta_k \hat G \eta_k + S[A_{2,3}]}. 
\end{align}
Here, $U$ is the diagonal matrix introduced after Eq.~\eqref{eq:Ht}.
Intersite correlator (obtained by differentiation with respect to $\eta$) thus contain averages like the following 
\begin{widetext}
\begin{eqnarray}
\langle e^{i A_{m}(\tau)}  e^{-i A_{m'}(0)} \rangle  &=& \tr[ e^{- (\beta - \tau) H} e^{-i A_m} e^{- \tau H} e^{i A_{m'}} ] \notag \\
& \stackrel{T\rightarrow 0}{=}& \sum_{\v p}\int d^2 x \int d^2 x' \psi^*_0( \vec x)e^{-i A_{m}( \vec x)}\psi_{\vec p}( \vec x) e^{-i\tau \epsilon_{\vec p}} \psi_{\vec p}^*( \vec x')e^{i A_{m'}(\vec x')}\psi_0(\vec x') \notag \\
&=& \delta_{mm'} e^{- i \epsilon_{-2\hat e_m/3}}.
\end{eqnarray}
\end{widetext}
We used the notation $\psi_{\vec p}( \vec x)$ for eigenstates of $(- i \nabla_{\vec x})^2/(2m_x)$ with eigenenergy $\vec p^2/(2m_x)$.
Thus, only intrasite terms survive.

\subsection{Order\red{ing} of $\mathcal O_s = d_{abc} \hat S^a_1 \hat S^b_2 \hat S^c_3$.}

For $t>t_c$ (i.e.~in the 3CK phase), the effective action of $\Phi$ fluctuations can be obtained by expansion about the minimum of $\cos(\Phi)$ leading to 
\begin{eqnarray}
S[\Phi] = \int \frac{d \omega}{2\pi} \Phi(\omega) \Phi(-\omega) \left [\frac{\eta}{4\pi} \vert \omega \vert  + \frac{M_\Phi}{2}\right ]
\end{eqnarray}
where $M_\Phi = N \beta T_K$. The correlator of phase fluctuations thus decays as
\begin{equation}
\langle \Phi(\tau) \Phi(0) \rangle \sim  \int_0^\infty d\omega \frac{\cos(\omega \tau)}{ \eta \vert \omega \vert + 2\pi M_\Phi} \sim -\frac{\eta \sin(2\pi M_\Phi \tau)}{(M_\Phi \tau)^2},
\end{equation}
and therefore leads to long-range correlations
\begin{equation}
\langle \mathcal O_s (\tau) \mathcal O_s (0) \rangle \sim  \frac{t^6}{T_{\rm K}^6} e^{-\frac{\eta \sin(2\pi M_\Phi \tau)}{(M_\Phi \tau)^2}} \rightarrow \frac{t^6}{T_{\rm K}^6}.
\end{equation}

\section{Phase slips}

Here we include phase slips of weight $\Gamma$  and time $\tau_0$ which is assumed to be smaller than all other time scales of the effective theory. We now consider a single kink in $\Phi$ with shift $2\pi$, which is associated to an amplitude~\cite{VainshteinShifman1982}
\begin{equation}
\mathcal A^{(0)}_{(-\tau/2, \Phi) \rightarrow (\tau/2, \Phi \pm 2\pi)} = \Gamma \int_{-\tau/2}^{\tau/2}  d \tau_c.
\end{equation}

\subsection{Instanton interactions}
\label{app:InstIA}

We consider the full partition function (generating functional) to second order in  $\Gamma$. 
\begin{equation}
Z[\eta] =  Z_0[\eta]+ \Gamma^2 \sum_\pm \int_0^\beta d\tau_f \int_{0}^{\tau_i} d\tau_i Z_{2,\pm}[\eta; \tau_f, \tau_i],
\end{equation}
where a phaseslip (anti phase slip ) is introduce at $\tau_i$ ($\tau_f$) and the sum over $\pm$ indicates the direction of the slip. The partition function is
\begin{align}
F &= - T  \ln Z[0] \notag \\
&\simeq \underbrace{-T \ln Z_0[0]}_{=F_{0}} - T \Gamma^2 \sum_\pm \int_0^\beta d\tau_f \int_{0}^{\tau_i} d\tau_i \frac{Z_{2,\pm}[0; \tau_f, \tau_i]}{Z_0[0]}.
\end{align}

We use that, before and after a phase slip, $h$ labels the same quantum states, however their energy has been shuffled around cyclically $\epsilon_h \rightarrow \epsilon_{h + 1}$. We can thus express the partition function in the helicity basis, $Z_0[\eta] = \prod_h Z_{0,h}[\eta_k]$, $Z_{2,\pm}[\eta; \tau_f, \tau_i] = \prod_h Z_{2,\pm,h}[\eta_h; \tau_f, \tau_i]$ where ($h$ index from now on suppressed unless explicitly restored)
\begin{subequations}\label{eq:2Slips}
\begin{align}
Z_0[\eta] &= \int \mathcal D [c,f] e^{- \int d \tau  (\bar c, \bar f) [\partial_\tau + H_{\rm MF}] (
c ,f )^T + \bar \eta f + \bar f \eta}\\
Z_2[\eta] &= \int \mathcal D [c,f] e^{- \int d \tau  (\bar c, \bar f) [\partial_\tau + H_{\rm slips}(\tau)] (
c ,f )^T + \bar \eta f + \bar f \eta},
\end{align}
and
\begin{equation}
H_{\rm slips} (\tau)  = \left (\begin{array}{cc}
\epsilon(\v p) & V \\ 
V & \lambda(\tau)
\end{array} \right )
\end{equation}
\end{subequations}
where $\lambda(\tau) = \lambda + \delta \lambda \chi_{\tau_i,\tau_f}(\tau)$ and $\chi_{\tau_i,\tau_f}(\tau) = 1$ for $\tau_i < \tau < \tau_f$ and $\chi_{\tau_i,\tau_f}(\tau) = 0$, otherwise. Note that $\lambda = \lambda_h = \lambda -\epsilon_h$, for the ground state $\delta \lambda = 3t$, for one of the excited state $\delta \lambda = -3t$ and for the third state $\delta \lambda = 0$ ($k = \pm 2\pi/3$ are degenerate).

Thus, the instantons generate an x-ray edge problem in each helicity channel. We follow~\cite{Gogolin2004} and employ the long-time f-electron Green's function
\begin{equation}
G_f (\tau ) \approx -\frac{g}{\tau},
\end{equation}
with $g = \Delta/\pi(\Delta^2 + \lambda^2)$ for the Kondo/resonant level problem. This leads to
\begin{equation}\label{eq:SSlips}
S_{\rm slips}(\tau_f - \tau_i) = (\tau_f - \tau_i) \delta \lambda {G_f(0,0^+)} {+} \left (\frac{\delta_x}{\pi}\right ) ^2 \ln \left (  \frac{\tau_f - \tau_i}{\lambda}\right),
\end{equation}
where $\delta_x = - \arctan(\pi g \delta \lambda)$. The first (classical) term cancels upon taking the product of $h$, leaving only the logarithmic repulsion. This concludes the derivation of $\kappa = 2 N (\delta_x/\pi)^2$, \red{Eq.~\eqref{eq:Kappa}}, as presented in the main text.

\subsection{Infinite order resummation of phase slips}
\label{app:InfResumSlips}

We now switch to the full resummation of phase slips. 
We consider an amplitude for $\Phi \rightarrow \Phi + N_\Phi 2\pi$ and denote $n$ and $\bar n$ the number of kinks/antikinks (i.e.~$n - \bar n = N_{\rm \Phi}$) and their center of mass time $\tau_1, \dots \tau_{n + \bar n}$. Different instanton sequences correspond to the integral over these variables. Then the amplitude is
\begin{widetext}
\begin{eqnarray}
\mathcal A_{(0, \Phi) \rightarrow (\beta, \Phi + N_\Phi 2\pi)} &=& \sum_{n = 0}^\infty \sum_{\bar n = 0}^\infty \delta_{n - \bar n, N_{\rm \Phi}}\; \Gamma^{n + \bar n} \sum_{\substack{\text{perm. of }\\ \bar n, n \text{ kinks}}} \Big \lbrace \int_{0}^{\beta}  d \tau_{\rm n + \bar n} \, \dots \int_{0}^{\tau_3}  d \tau_{2} \int_{0}^{\tau_2}  d \tau_{1} \notag  \\
&&\tr[e^{ - \int_{\tau_{n + \bar n}}^{\beta} d\tau' \hat H_{\rm MF}(\Phi+N_\Phi 2\pi)} \dots e^{ - \int_{\tau_{1}}^{\tau_2} d\tau' \hat H_{\rm MF}(\Phi\pm 2\pi)}e^{ - \int_{\tau_{0}}^{\tau_1} d\tau' \hat H_{\rm MF}(\Phi)}] \Big \rbrace. \label{eq:InstGas}
\end{eqnarray}
\end{widetext}
In the second line, the $\pm$ refers to the sign of the first kink. We can now use that the Hamiltonian between two kinks is time independent and the evolution operator between two kinks is 
\begin{align}
\mathcal T e^{-\int_{\tau_l}^{\tau_{l+1}} d\tau' \hat H_{\rm MF}(\Phi+k 2\pi)}&= \prod_{\Delta \tau} \left [1 + \Delta \tau \hat H(\Phi + k 2\pi) \right ]\notag \\
&=\prod_{\Delta \tau} \left (\tau_\Phi^k \left [1 + \Delta \tau H(\Phi) \right ] \tau_\Phi^{-k} \right)\notag\\
&=\tau_\Phi^k \prod_{\Delta \tau} \left ( \left [1 + \Delta \tau H(\Phi) \right ]  \right)\tau_\Phi^{-k}\notag\\
&=\tau_\Phi^k \mathcal T [e^{-\int_{\tau_l}^{\tau_{l+1}} d\tau' H_{\rm MF}(\Phi)}] \tau_\Phi^{-k}.
\end{align}
Thus, a kink at time $\tau$ is represented by the operator insertion $\tau_\Phi$ at time $\tau$ into the partition sum

\begin{widetext}
\begin{align}
\mathcal A_{(0, \Phi) \rightarrow (\beta, \Phi + N_\Phi 2\pi)} &= \sum_{n = 0}^\infty \sum_{\bar n = 0}^\infty \delta_{n - \bar n, N_{\rm \Phi}}\; \Gamma^{n + \bar n}  {\frac{(n + \bar n)!}{n! \bar n!}} {\frac{1}{(n+\bar n)!}} \notag \\
&\mathcal T \Big \lbrace \int_0^\beta  d \tau_{\rm n} \dots \int_0^\beta  d \tau_{1} \int_0^\beta d \bar \tau_{\rm \bar n} \dots \int_0^\beta  d \bar \tau_{1} \tr[\prod_{j = 1}^n\prod_{\bar j = 1}^{\bar n} \tau_\Phi (\tau_k) \tau_\Phi^{-1} (\bar\tau_{\bar j}) e^{- \int_0^\beta d\tau \hat H_{\rm MF}}] \Big \rbrace \notag \\
&= \frac{1}{3}\sum_\theta e^{-i \theta N_{\Phi}}\tr [e^{\int_{0}^\beta d\tau \Gamma e^{i \theta} \tau_\Phi (\tau)}e^{\int_{0}^\beta d\tau \Gamma e^{-i \theta} \tau_\Phi^{-1} (\tau)}e^{-\int_{0}^\beta d\tau \hat H_{\rm MF}}].
\end{align}
\end{widetext}

We used the Fourier transform on $\ell^3$ with periodic boundary conditions (i.e.~with possible wavevectors $\theta = 0, \pm 2\pi/3$) such that $\sum_\theta e^{i \theta N} = 3\delta_{N,0}$ and $\sum_n e^{i n \theta} = 3 \delta_{\theta,0}$. 
Here, ${\frac{(n + \bar n)!}{n! \bar n!}}$ is the number of possibilities to arrange the $n$ upsteps if there are $n+ \bar n$ steps in total. The factor ${\frac{1}{(n+\bar n)!}}$ accounts for the fact that the integration domain has been increased from an explicitly time ordered $n+\bar n$ dimensional integral in Eq.~\eqref{eq:InstGas}, to a $n+\bar n$ dimensional hypercube.

The total partition sum is given by (we use $\sum_{N_{\Phi}} e^{-i N_{\Phi} \theta} = 3 \delta_{\theta,0}$ for $N_\Phi = 0, \pm 1$)
\begin{equation}
Z = \sum_{N_\phi} \mathcal A_{(0, \Phi) \rightarrow (\beta, \Phi + N_\Phi 2\pi)} = \tr[e^{- \beta [\hat H_{\rm MF} - \Gamma (\tau_\Phi + \tau_\Phi^{-1})]}] 
\end{equation}

We now restore the matrix space of different vacua. In total we obtain

\begin{eqnarray}
H_{\rm eff}&=& \sum_{x} [-t_c c^\dagger_{\alpha m}(x)  c_{\alpha m} (x+1) + h.c. - \mu c^\dagger_{\alpha m}(x)  c_{\alpha m} (x)] \mathbf 1_\Phi \notag\\
&&+ [t \sigma_\Phi f^\dagger_{\alpha,m} f_{\alpha, m+1} + h.c.] + \lambda f^\dagger_{\alpha m}f_{\alpha m} \mathbf 1_\Phi \notag \\
&&+ [V f^\dagger_{\alpha,m} c_{\alpha, m} + h.c.] \mathbf 1_\Phi- \Gamma (\tau_\Phi+\tau_\Phi^{-1}) . 
\end{eqnarray}

\red{This concludes the derivation of Eq.~\eqref{eq:Heff} of the main text.}

\subsection{Orthogonality catastrophe using bosonization}
We start from the effective Hamiltonian derived in the previous section
\be
H=H_0+H_\Gamma, \qquad H_\Gamma=-\Gamma\hat O, \qquad \hat O=\tau_\Phi\dn+\tau_\Phi^{-1}.
\ee
We are going to treat this problem perturbatively to second order in $\Gamma$ and diagonalize $H_0$ in the helicity $h$ basis. Then, considering that $\lambda_h[\Phi]$ is different for $\Phi=-2\pi,0,2\pi$ we have 
\begin{align}
H_0 &=\sum_{\Phi=0}\sum_h\ket{\Phi}H_{0h}[\Phi]\bra{\Phi},\notag \\
H_{0h}[\Phi]&=
\mat{c\\ f}_h\dg
\mat{\eps_c & V \\ V & \lambda_h[\Phi]}\mat{c\\ f}_h.
\end{align}
Note that $c$-electrons have another momentum $k$ along the wires, which is implicit here. This problem as is, is difficult to treat. We are forced to i) go to the scattering basis $\psi_{h\sigma}$ and ii) assume that the phase shift is independent of the energy, i.e.~the electrons in scattering basis experience a potential scattering $\tilde V_h[\Phi]$ which depends on the flux $\Phi$. In that case we can unfold the conduction electrons to right-movers only and write
\begin{align}
H_{0h}[\Phi]&=H_{0h}+\sqrt{2\pi}\tilde V_h[\Phi]\psi\dg_{h\sigma}(0)\psi\dn_{h\sigma}(0), \notag \\
 H_{0h} &=-iv_F\int{dx\psi\dg_{h\sigma}\partial_x\psi\dn_{h\sigma}}
\end{align}
where the relation between the potential scattering $\tilde V_h[\Phi]$ and the phase shift is shown below in Eq.\,(\ref{eqs111}) and the factor of $\sqrt{2\pi}$ is introduced for future convenience. Next, we bosonize, i.e.~express the fermions as
\be
\psi_h(x)\sim e^{i\sqrt{2\pi}\varphi_h(x)}, \; [\varphi_h(x),\varphi_{h'}(y)]=\frac{i}{2}\sgn{x-y}\delta_{hh'}.
\ee
The Hamiltonian becomes
\be
H_{0h}[\Phi]=H_{0h}+(\partial_x\varphi_h)\tilde V_h[\Phi],
\; H_{0h}=\frac{v_F}{2}\int_0^\infty{dx}(\partial_x\varphi_h)^2.
\ee
It is easy to see that the potential scattering term can be eliminated
\begin{align}
H_{0h}[\Phi]&\equiv\frac{v_F}{2}\int{dx}\Big\{\partial_x\varphi_h+\delta(x)\tilde V_h[\Phi]/v_F\Big\}^2, \notag \\
& \text{i.e.}\quad \varphi_h(x)\to \varphi_h(x)+\theta(x)\tilde V_h[\Phi]/v_F.
\end{align}
By plugging this into $\psi\sim e^{i\sqrt{2\pi}\varphi}$ we can see that this corresponds to the phase shift 
\be
\psi_{out,h}=\psi_{in,h} e^{2i\delta_h}, \qquad \delta_h=\sqrt{\frac{\pi}{2}}\frac{\tilde V_h}{v_F}\label{eqs111}
\ee
Each flux configuration corresponds to a different phase shift in a given helicity sector and these configurations are related to each other via the so-called Schotte-Schotte transformation \cite{Affleck1994c}
\be
U_h[\Phi,\Delta\Phi]=\exp\Big\{-i\varphi(0)\Big(\tilde V_h[\Phi+\Delta\Phi]-\tilde V_h[\Phi]\Big)/v_F\Big\}.
\ee
Using the commutation relation of bosons and the fact that $e^{sX}Ye^{-sX}=Y+s[X,Y]$, for $[X,Y]$ c-number, we can check that
\bea
H_{0h}[\Phi+\Delta\Phi]&=&U_h\dg[\Phi,\Delta\Phi] H_{0h}[\Phi]U_h[\Phi,\Delta\Phi]\label{eqs113}\\
&=&H_{0h}+(\partial_x\varphi)\tilde V_h[\Phi]\notag\\
&-&i[\tilde V(\Phi+\Delta\Phi)-\tilde V(\Phi)]\int_x\partial_x\varphi[\varphi(0),\partial_x\varphi] \notag\\
&=&H_{0h}+{\tilde V}[\Phi+\Delta\Phi](\partial_x\varphi)\vert_{x=0}.
\eea
Going to interaction picture w.r.t. $H_0$ and expanding the partition function in $\Gamma$ we have
\bea
Z/Z_0&=&\braket{T_\tau e^{-\int_{-1/2T}^{1/2T}{d\tau}H_{\Gamma}(\tau)}}_0 \notag \\
&=&1+\frac{\Gamma^2}{2}\int{d\tau_1d\tau_2}\braket{T_\tau e^{\tau_1H_0}\hat O e^{(\tau_2-\tau_1)H_0}\hat Oe^{-\tau_2H_0}}_0\notag \\
&=&1+\frac{\Gamma^2}{2}\int{d\tau_1d\tau_2}\braket{T_\tau e^{(\tau_1-\tau_2)H_0} e^{(\tau_2-\tau_1)\hat O H_0\hat O}}_0\notag\\
&=&1+\frac{\Gamma^2}{2}\sum_{\Phi}\sum_{\alpha=\pm1}\int{d\tau_1d\tau_2}\prod_h\notag \\
&&\times\braket{T_\tau e^{(\tau_1-\tau_2)H_{0h}[\Phi]} e^{(\tau_2-\tau_1)H_{0h}[\Phi+2\pi\alpha]}}_0
\eea
where $\hat O=\tau^{+1}+\tau^{-1}$, we used that the linear-in-$\Gamma$ term vanishes due to trace and use the cyclic property of the trace with the Boltzmann factor $e^{-\beta H_0}/Z_0$ to shuffle the time-evolutions. Using Eq.\,(\ref{eqs113}):
\be
e^{\tau H_{0h}[\Phi+\Delta\Phi]}=U\dg_h[\Phi,\Delta\Phi]e^{\tau H_{0h}[\Phi]}U_h[\Phi,\Delta\Phi]
\ee
we can write
\bea
&&\braket{T_\tau e^{-\Delta\tau H_{0h}[\Phi]}U\dg_h[\Phi,\Delta\Phi]e^{\tau H_{0h}[\Phi]}U_h[\Phi,\Delta\Phi]}_0\notag \\
&=&\braket{T_\tau U\dg_h[\Phi,\Delta\Phi;\tau]U_h[\Phi,\Delta\Phi]}_0\\
&=&\braket{T_\tau e^{i\varphi(\tau)\Delta V_h[\Phi,\Delta\Phi]/v_F}e^{-i\varphi(0)\Delta V_h[\Phi,\Delta\Phi]/v_F}}\notag \\
&\sim& \abs{\tau}^{-\frac{(\Delta \tilde V_h[\Phi,\Delta\Phi])^2}{2\pi v_F^2}}\nonumber
\eea
where we used that
\be
\braket{e^{i\gamma\varphi(\tau)}e^{-i\gamma\varphi(0)}}=\frac{1}{\abs{\tau}^{\gamma^2/2\pi}}.
\ee
As a reminder, $\Delta \tilde V_h$ can be related to the phase shift
\begin{align}
\Delta \tilde V_h[\Phi,\Delta\Phi]&\equiv \tilde V_h[\Phi+\Delta\Phi]-\tilde V_h[\Phi]\notag \\
&=v_F\sqrt{\frac{2}{\pi}}\Big(\delta_h[\Phi+\Delta\Phi]-\delta_h[\Phi]\Big)
\end{align}
which leads to
\bea
Z/Z_0&=&1+\frac{1}{T}{\Gamma^2}\sum_\Phi\sum_{\alpha=\pm 1}\int_{-1/2T}^{1/2T} d\Delta\tau\abs{\Delta\tau}^{-\kappa }, \notag\\
 \kappa&=&\sum_h(\delta_h[\Phi+\Delta\Phi]/\pi-\delta_h[\Phi]/\pi)^2.
\eea
Up to subleading terms in small $t$ (which are not important near the transition), this exactly reproduces Eq.~\eqref{eq:SSlips}. The time-integral leads to
\be
Z/Z_0=1+{\cal C}'{\Gamma^2}T^{\kappa-2}
\ee
where ${\cal C}'$ is a constant. The corrections to free energy $F_0=-T\log Z_0$ is 
\be
F-F_0=-{\cal C}'\Gamma^2T^{\kappa-1}.
\ee
\subsection{$\braket{\sigma_\Phi(\tau)\sigma_\Phi(0)}$ correlation function}
In this section we compute the correlator $\braket{\sigma_\Phi(\tau)\sigma_\Phi(0)}$ which is related to the order parameter $\braket{{\cal O}_s(\tau){\cal O}_s(0)}$ or $\braket{\Phi(\tau)\Phi(\tau)}$ in the paper, within the $t-\Gamma$ Hamiltonian $H=H_0+H_\Gamma$. In the $\Gamma/t\gg 1$ regime (FL phase) this is exponentially decaying. This can be seen easily in a basis in which the $-\Gamma O$ term is diagonal.
In the limit of large $\Gamma$, we can use a unitary transformation $U_O$ to diagonalize $O$ 
\be
O=\tau\dn_\Phi+\tau^{-1}_\Phi=\matc{ccc}{0 & 1 & 1 \\ 1 & 0 & 1 \\ 1 & 1 & 0} \to U_O\dg OU\dn_O=\matc{ccc}{-1 \\ & -1 \\ && 2}
\ee
and go to the interaction picture w.r.t. $-\Gamma\tau^x$. In this picture $\sigma(\tau)$ is time-dependent and is given by
\begin{align}
\rho_\Gamma&=\frac{e^{-\Gamma/T}}{2e^{-\Gamma/T}+e^{2\Gamma/T}}\matc{ccc}{1\\ & 1 \\ && e^{3\Gamma/T}}\to\matc{ccc}{0\\ &0\\ && 1}
,
\notag\\
\sigma_\Gamma(\tau)&=\frac{\omega^2}{2}\matc{ccc}{1 & i & -x^{-1}\sqrt 2 \\ i & -1 & ix^{-1}\sqrt 2 \\ -x\sqrt{2} & ix\sqrt{2} & 0},
\end{align}
in terms of $x=e^{3\tau\Gamma}$. With this density matrix ${\rm tr}[\rho_\Gamma O]=o_{33}$. Therefore to leading order in tunnelling $t$,
\begin{align}
\braket{\sigma_\Gamma}&=\braket{T_\tau\sigma_\Gamma(\tau)\sigma_\Gamma(0)}=0, \notag\\
 \braket{T_\tau\sigma_\Gamma(\tau)\sigma\dg_\Gamma(0)}&=\braket{T_\tau\sigma_\Gamma\dg(\tau)\sigma_\Gamma(0)}=e^{-3\abs{\tau}\Gamma}.
\end{align}
\red{T}his is the origin of the fact that $t$ is irrelevant in the $\Gamma/t\gg 1$ regime, within $t-\Gamma$ Hamiltonian. In the opposite regime of $\Gamma/t\ll 1$, we can use the same technique as in previous section to compute the Since $\sigma_\Phi$ commutes with $H_0$, to zero order in $\Gamma$ we have
\be
\braket{\sigma\dn_\Phi(\tau)\sigma\dg_\phi(0)}=\frac{1}{3}\tr[{\sigma\dn_\Phi\sigma\dg_\phi}]=1.
\ee
To second order in $\Gamma$ we have (we have neglected the disconnected part, since it does not depend on $\tau$)
\begin{align}
&\braket{\sigma\dn_\Phi(\tau)\sigma\dg_\phi(0)}=\notag\\
&1+\Gamma^2\int_{-1/2T}^{1/2T}d\tau_1d\tau_2\braket{T_\tau \sigma\dn_\Phi(\tau)\sigma\dg_\Phi(0) \hat O(\tau_1) \hat O(\tau_2)}
\end{align}
We can divide the integration range into 6 configurations (assuming $\tau_1>\tau_2$):
\begin{widetext}
\bea
&&\theta_1\equiv \theta(\tau_1>\tau_2>
\tau>0):\qquad \sum_{\alpha\alpha'}\braket{e^{\tau_1H_0}\tau_\Phi^\alpha e^{(\tau_2-\tau_1)H_0}\tau_\Phi^{\alpha'}e^{-\tau_2H_0}\sigma\dn_\Phi \sigma\dg_\Phi}= G(\Delta\tau)\\
&&\theta_2\equiv \theta(\tau>0>
\tau_1>\tau_2):\qquad \sum_{\alpha\alpha'}\braket{\sigma\dn_\Phi\sigma\dg_\Phi e^{\tau_1H_0}\tau_\Phi^\alpha e^{(\tau_2-\tau_1)H_0}\tau_\Phi^{\alpha'}e^{-\tau_2H_0}}=\omega G(\Delta\tau)\\
&&\theta_3\equiv \theta(\tau>
\tau_1>\tau_2>0):\qquad \sum_{\alpha\alpha'}\braket{\sigma\dn_\Phi e^{\tau_1 H_0}\tau_\Phi^\alpha e^{(\tau_2-\tau_1)H_0}\tau_{\Phi}^{\alpha'}e^{-\tau_2H_0}\sigma\dg_\Phi}=G(\Delta\tau)\\
&&\theta_4\equiv \theta(\tau_1>\tau>
0>\tau_2):\qquad \sum_{\alpha\alpha'}\braket{e^{\tau_1H_0}\tau_\Phi^\alpha e^{-\tau_1H_0}\sigma\dn_\Phi\sigma\dg_\Phi e^{\tau_2H_0} \tau_\Phi^{\alpha'}e^{-\tau_2H_0}}=\bar\omega G(\Delta\tau)\\
&&\theta_5\equiv \theta(\tau>\tau_1>0>\tau_2):\qquad \sum_{\alpha\alpha'}\braket{\sigma\dn_\Phi e^{\tau_1H_0}\tau_\Phi^\alpha e^{-\tau_1 H_0}\sigma\dg_\Phi e^{\tau_2H_0}\tau_\Phi^{\alpha'}e^{-\tau_2H_0}}=G(\Delta\tau)\\
&&\theta_6\equiv \theta(\tau_1>\tau>\tau_2>0):\qquad \sum_{\alpha\alpha'}\braket{e^{\tau_1H_0}\tau^\alpha_\Phi e^{-\tau_1H_0}\sigma\dn_\Phi e^{\tau_2H_0}\tau_\Phi^{\alpha'}e^{-\tau_2H_0}\sigma\dg_\Phi}=G(\Delta\tau)
\eea
\end{widetext}
Here, $\alpha,\alpha'=+1,-1$ and we have used the $\sigma_\Phi\tau_\Phi=\omega\tau_\Phi\sigma_\Phi$ and similar commutation relations to eliminate $\sigma_\Phi$ and $\tau_\Phi$ and express the correlators in terms of a single correlator ($\Delta\tau\equiv \tau_1-\tau_2$)
\be
G(\Delta\tau)=\sum_\alpha\sum_\Phi\prod_h\braket{e^{\Delta\tau H_0[\Phi]}e^{-\Delta\tau H_0[\Phi+2\pi\alpha]}},
\ee
which is the correlator was computed in the previous section
The integration over these ranges appear with a integrand that is only a function of $\tau_1-\tau_2$. Denoting,
\be
I_i\equiv\int{d\tau_1d\tau_2\theta_i}G(\tau_1-\tau_2)
\ee
we have
typical integrals of the form 
\be
I\sim \tau\int_0^\tau d \Delta \tau G(\Delta\tau), \qquad G(\Delta\tau)\sim \vert \Delta\tau \vert ^{-\kappa}
\ee
in terms of $\kappa$ defined before, which gives us
\be
\braket{\sigma_\Phi(\tau)\sigma_\Phi^\dagger(0)}\sim 1+{\cal C}''\Gamma^2\tau^{2-\kappa}.
\ee
where ${\cal C}''$ is another constant. This demonstrates that the $\braket{\sigma_\Phi(\tau)\sigma_\Phi(0)}$ correlator 
disorders at the deconfinement quantum phase transition, defined by $\kappa = 2$. 
%


\bibliography{FrustratedTriangle}
\end{document}